\documentstyle[aps,prd,amsfonts,amssymb,epsfig,floats]{revtex}

\def\aa_{Astron.~Astrophys.}

\def\app{Astropart.~Phys.~}
\def\apj{Astrophys.~J.~}

\def\epjc{Eur.~Phys.~J.~C~}
\def\mnras{Mon.~Not.~R.~Astron.~Soc.~}
\def\npa{Nucl.~Phys.~A~}

\def\plb{Phys.~Lett.~B~}
\def\rmp{Rev.~Mod.~Phys.~}
\def\ijmpa{Int.~J.~Mod.~Phys.~A~}
\def\jmp{J.~Math.~Phys.~}

\newcommand{\rbox}[1]{\raisebox{1.5ex}[0pt]{#1}}

\draft    

\tighten
\begin{document}
\twocolumn[\hsize\textwidth\columnwidth\hsize\csname
@twocolumnfalse\endcsname
\widetext

\title{Can three-flavor oscillations solve the solar
neutrino problem?}

\author{H. Schlattl\footnotemark[1]}
\address{Max-Planck-Institut f\"ur Astrophysik, Karl-Schwarzschild-Str. 1, 
85741 Garching, Germany}
\date{\today}

\maketitle

\begin{abstract}
The good agreement of standard solar models with helioseismology and
the combined analysis of the solar neutrino experiments suggest that
the solution to the solar neutrino problem is located in particle
physics rather than in astrophysics. The most promising solution are neutrino
oscillations, which usually are analyzed within the reduced 2-flavor
scheme, because the solutions found therein reasonably
well reproduce the recent data of Super-Kamiokande about the 
recoil-electron energy spectrum, zenith-angle and seasonal variations,
and the event rate data of all the neutrino detectors. In this work,
however, a survey of the complete parameter space of 3-flavor
oscillations is performed. Basically
eight new additional solutions could be identified, where the
best one with $\Delta m^2_{12}=2.7\times
10^{-10}\,{\rm eV^2}$, $\Delta m_{13}^2=1.0\times10^{-5}\,{\rm eV^2}$,
$\Theta_{12}=23^\circ$, and $\Theta_{13}=1.3^\circ$ (denoted SVO)
is slightly more probable than any 2-flavor solution.
While the 2-flavor results of the 
atmospheric neutrino problem ($\Delta m_{23}^2\gtrsim10^{-3}\,{\rm
eV^2}$) would exclude all the 
3-flavor solutions of this work, in the 3-flavor atmospheric-neutrino
analysis the SLMA-solution with $\Delta m^2_{12}=7.9\times 10^{-6}\,{\rm
eV^2}$, $\Delta m_{13}^2=2.5\times10^{-4}\,{\rm eV^2}$,
$\Theta_{12}=1.4^\circ$, and $\Theta_{13}=20^\circ$ is still allowed.
The relatively weak
improvement of the fit using 3-flavor instead of 2-flavor
oscillations, which appears to be due to an inconsistency of the
different kind of data, indicates that there are possibly still systematic
errors in at least one data set, or that the statistics is not yet
sufficient. Besides, the ability of SNO and
Borexino to discriminate the various 2- and 3-flavor solutions
is investigated. Only with very good statistics in these experiments 
the correct solution to the solar neutrino problem can be identified
unambiguously.
\end{abstract}
\vspace*{.3cm}]
\footnotetext[1]{e-mail address: schlattl@mpa-garching.mpg.de}
\stepcounter{footnote}
\narrowtext

\section{Introduction}
{From} the beginning of the first measurements of the solar neutrino
flux on Earth~\cite{Dav68} until the present time, the origin of the
solar neutrino problem could not yet be resolved totally. While
previously inaccurate or unknown physics used in solar-model
calculations could have been made responsible for the discrepancy between
measured and predicted solar neutrino flux, this kind of solution can
presently almost be ruled out, basically of two reasons:

Firstly, the high precision in the measurements of the solar p-mode
frequencies and the development of helioseismological inversion
techniques enable the determination of the solar sound-speed profile
with high accuracy~\cite{pmodel}. The comparison with
standard solar models containing improved input physics like opacity,
equation of state and microscopic diffusion shows an excellent
agreement with the seismic
models~\cite{SWL97,Bah98,Gab96}. Predictions for the event
rates in the solar neutrino detectors deduced from these standard
solar models still are inconsistent with the measurements, confirming
the solar neutrino problem (Table~\ref{neurate}).

Secondly, the three types of currently operating experiments, 
the chlorine detector~\cite{Dav68}, the gallium experiments
GALLEX/GNO\cite{GALLEX92} and SAGE~\cite{SAGE94}, and the \v{C}erenkov-light
counter Super-Kamiokande have different neutrino-energy
thresholds. This allows to determine the contribution of different
parts of the solar neutrino spectrum to the total flux without
explicitly taking into account solar-model calculations. {From}
this analysis it has been inferred that the experimental results can
be explained only with huge changes in the nuclear
fusion rates. The best fit with the data is obtained 
even with a \emph{negative} flux of neutrinos created in the
electron-capture process of $^7$Be~\cite{Hat95}. Nevertheless, strong
modifications of 
the reaction cross sections would be difficult to explain
experimentally and theoretically. Moreover, even if a presently
unknown physical process can account for the demanded changes, the
resulting solar models would hardly be consistent with
helioseismology~\cite{PhD}. 

The most promising approach to the solution of the solar neutrino
problem is an extension to the particle-physics standard model ---
neutrino mixing. Analogous to the CKM-mixing in the quark sector, weak
and mass eigenstates of the neutrinos are supposed not to be 
identical but connected by a unitary transformation. 

Under this assumption an initial solar electron-neutrino can be
converted during its propagation to the Earth into another flavor, a
$\mu$- or $\tau$-neutrino (\emph{just-so}
oscillations~\cite{Pon67}). 
Furthermore,  the neutrinos may coherently scatter forward in
solar matter (\emph{MSW-effect}~\cite{MS}) altering the
conversion probability for a certain set of  
mixing parameters.  

The possible values for the mixing parameters, with which the measured
event rates in all detectors can be reproduced simultaneously, have
been derived by various authors~\cite{Hat97,BahK97}, but
only the oscillations between two flavors usually is taken into
account. 

Recently Super-Kamiokande has published more detailed information
about the energy distribution of the recoiled electrons and the
zenith-angle dependence of the neutrino-signal~\cite{SK99e,SK99z}. 
In the analysis of the 825-days data it became clear, that it is not
possible to explain satisfactorily these data \emph{and} the event
rates of all detectors by 
one set of mixing parameters (see \cite{BahK98}):  
An excess of event rates in the high-energy bins 
was inconsistent with the other data. 

It was the initial motivation of the present work to examine, whether
an expansion of the neutrino analysis to the more general
3-flavor case could resolve this discrepancy between the
different types of data.
However, after the Super-Kamiokande group reanalyzed their data and included
new data (1117 days in total), in particular the excess in the
energy-bin data could be diminished and now all kind of data can be
explained simultaneously by 2-flavor oscillations. Furthermore,
the neutrino spectrum in the $^8$B-decay has been measured recently
in the laboratory~\cite{OGW00}. Although within the errors the
spectrum is in agreement with the one predicted theoretically
in~\cite{Bah96}, 
the number of high-energetic neutrinos is overall higher than
previously thought. Thus, the excess in the high-energy bins
is further 
reduced which yields a yet slightly better reproduction of the data by
the 2-flavor solutions (see section \ref{twoflav}).
Nevertheless, it
is presently still not clear, which solution to the solar neutrino
problem is the correct one, and hence all possible solution should be
deduced.

Therefore, in this work the most general case of 3-flavor oscillations
is investigated without making any assumptions about the mass scale
as, for instance, inspired by the atmospheric neutrino problem. The
latter is taken in various publications \cite{Fog96,Fog00,TSI99} as a
constraint to investigate 3-flavor oscillations. 
But here, the aim is to examine whether the expansion to three
flavors leads to new solutions with which the fits to all
kind of data can be improved compared to the usual 2-flavor analysis.
The implications for the atmospheric neutrino puzzle are discussed
afterwards.  

In section~\ref{theory} the equations for 3-flavor oscillations
are derived from the solution to the Klein-Gordon equation and the
size of the parameter space is deduced. After describing the
underlying solar model and the neutrino analysis in
section~\ref{calc}, the results for 2-flavor and 3-flavor
oscillations are shown (section~\ref{results}). Finally the ability of
forthcoming experiments like SNO and Borexino to discriminate the
various solutions are discussed (section~\ref{future}). 

\section{Theory of neutrino oscillations}\label{theory}
In the following an overview of the basic equations for neutrino
oscillations are provided with particular emphasis to the
3-flavor case. A more thorough description can be found e.g.~in
\cite{Raf96} or \cite{PhD}. 

\subsection{Vacuum oscillations}
If neutrinos have mass, a mixing matrix similar to the
Kobayashi-Maskawa matrix in the quark sector can be developed 
\begin{equation}\label{eigendef} 
|\nu_\alpha\rangle\; = \;\left(\begin{array}{c} {\nu_e}\\{\nu_\mu}\\{\nu_\tau}\end{array}\right)\; =\;
{\cal U}\,
\left(\begin{array}{c} {\nu_1}\\{\nu_2}\\{\nu_3}\end{array}\right)\;=\;{\cal U}\,|\nu_i\rangle,
\end{equation}
where $|\nu_\alpha\rangle$ denotes the weak
and $|\nu_i\rangle$ the mass eigenstates.
The unitary matrix ${\cal U}$ can be parameterized by 
\begin{eqnarray} \nonumber
{\cal U} & = & \left(\begin{array}{ccc}
1 & 0 & 0 \\
0 & c_{23} & s_{23} \\
0 & -s_{23} & c_{23} \\
\end{array}\right) \left(\begin{array}{ccc}
c_{13} & 0 & s_{13}e^{-i\delta} \\
0 & 1 & 0 \\
-s_{13}e^{i\delta} & 0 & c_{13} \\
\end{array}\right) \\\label{udef}
& & \times \left(\begin{array}{ccc}
c_{12} & s_{12} & 0 \\
-s_{12} & c_{12} & 0 \\
0 & 0 & 1 \\
\end{array}\right) 
\end{eqnarray}
where $s_{ij}$ and $c_{ij}$ are abbreviations for $\sin\Theta_{ij}$
resp.~$\cos\Theta_{ij}$ ($0\le\Theta_{ij}<\pi/2$) and $\delta$ is a
CP-violating phase~\cite{Pet88}, which is neglected in the
following\footnote{The effect of the CP-violating phase on the analysis of
neutrino oscillation data has been elaborated in \cite{DFL99}}.
The equation of motion for a neutrino beam in vacuum obeys
the Klein-Gordon equation for free particles ($\hbar=c=1$)
\begin{equation}\label{klein}
(\partial_{\rm t}^2 - \nabla^2 + M^2)\, |\nu_i(t,\vec{r})\rangle\; =
\;0,
\end{equation}
where the mass matrix $M^2$ is defined as 
\[M^2 = \left(\begin{array}{ccc}
m_1^2 & 0 & 0 \\
0 & m_2^2 & 0 \\
0 & 0 & m_3^2 
\end{array} \right)\]
with $m_i$ being the mass of the neutrino mass eigenstate $\nu_i$. Generally
the solution is given by a superposition of plane waves
$e^{i(\vec{k}\vec{r} - \omega t)}$ with the
dispersion relation
\[
\omega^2 = \vec{k}^2 + m^2. \]
In the case of the Sun with an almost stationary neutrino flux,
$|\nu_i(t,\vec{r})\rangle$ can be expanded in 
components of fixed frequency $|\nu_i(\vec{r})\rangle_{\!_\omega}
\,e^{-i\omega t}$. For a spherically symmetric flux of relativistic
neutrinos ($k\approx\omega$) one finally gets
\begin{equation}\label{evol}
-i\frac{\partial}{\partial r} |\nu_i(r)\rangle_{\!_\omega} =
 \left(\omega - \frac{M^2}{2\omega}\right)|\nu_i (r)\rangle_{\!_\omega} .
\end{equation}
The constant ``potential'' $\omega$ can be removed by shifting the energy
scale, and by using $t=r$ \mbox{Eq.~(\ref{evol})} can be formally written as
a more familiar Schr\"odinger-type equation 
\begin{equation} \label{schroe}
i\frac{\partial}{\partial t}|\nu_i(t)\rangle_{\!_\omega} = {\cal H}
|\nu_i(t)\rangle_{\!_\omega} 
\end{equation}
with ${\cal H}=E_{\rm kin}\approx
\frac{M^2}{2\omega}$. 
The general solution to this equation is 
\begin{equation} \label{vacsol}
|\nu_i(t)\rangle_{\!_\omega} = e^{-it \frac{M^2}{2\omega}}
 |\nu_i(t_0)\rangle_{\!_\omega}. 
\end{equation}
The probability $P(r)_{\alpha\rightarrow\beta}$ to detect a neutrino
$\nu_\alpha$ with energy $E=\omega$ as a neutrino of type
$\nu_\beta$ at distance $r=t$ 
from the source is therefore given by  (${\cal U}^{-1}={\cal U}^\dagger$)
\begin{equation} \label{prob}
P(r)_{\alpha \rightarrow \beta} =  
\left|\,\langle \nu_\beta |{\cal U} e^{-ir \left(\frac{\Delta M^2}{2E}\right)} 
 \cal{U}^\dagger|\nu_\alpha \rangle\, \right|^2.
\end{equation}
Using $\Delta m^2_{ij}=m_j^2-m_i^2$ the matrix $\Delta M^2$ writes as
\[ 
\Delta M^2 = M^2 - m_1^2\, I\!d_3 = \left(\begin{array}{ccc}
0 & 0 & 0 \\
0 & \Delta m^2_{12} & 0 \\
0 & 0 & \Delta m^2_{13} 
\end{array} \right).
\] 
With the mass eigenstate of an electron-neutrino being
\begin{equation}\label{electron}
\left(\begin{array}{c} {\nu_1}\\{\nu_2}\\{\nu_3}\end{array}\right)_{\!\!\rm e}\; =\; \cal{U}^\dagger
\,\left(\begin{array}{c} {1}\\{0}\\{0}\end{array}\right)\;=\;\left(
\begin{array}{c} {c_{12}c_{13}}\\{s_{12}c_{13}}\\{s_{13}} \end{array}\right),
\end{equation}
\mbox{Eq.~(\ref{prob})} yields for the survival probability of $\nu_e$'s in vacuum
\begin{eqnarray}\nonumber
P(r)_{e\rightarrow e}^{(3)} &  =  & 1 -
\sin^2\left(\!\frac{\Delta m^2_{12}\,r}{4E}\!\right)\, \cos^4\!\Theta_{13}\,
\sin^2\!2\Theta_{12}\\ \nonumber
& - & \Biggl[\sin^2\!\left(\!\frac{\Delta m^2_{13}\,r}{4E}\!\right)\,
\cos^2\!\Theta_{12} \\ \label{Pvac}
& & \quad + \sin^2\!\left(\!\frac{\Delta m^2_{23}\,r}{4E}\!\right) 
 \, \sin^2\!\Theta_{12} \Biggr]\;\sin^2\!2\Theta_{13}.
\end{eqnarray}
The superscript $(3)$ denotes
the case of 3-flavor mixing. $P_{e\rightarrow e}^{(3)}$ depends
on four quantities, two mass-squared differences,
$\Delta m^2_{12}$ and $\Delta m^2_{13}$ ($\Delta m^2_{23}=\Delta
m^2_{13} - \Delta m^2_{12}$), and two mixing angles
$\Theta_{12}$ and $\Theta_{13}$. The third mixing angle $\Theta_{23}$ does not appear in
\mbox{Eq.~(\ref{electron})} and hence $P_{e\rightarrow e}^{(3)}$ is independent of this 
quantity. The survival probabilities for $\nu_\mu$ and
$\nu_\tau$ depend on
$\Theta_{23}$, but its value cannot be determined by solar-neutrino
experiments, as 
in the energy range of the solar neutrinos $\nu_\mu$~and $\nu_\tau$~interact
equally with the detector material via NC-interactions. 
Thus, only the total number of $\mu$- plus $\tau$-neutrinos, given by
1$-$$P_{e\rightarrow e}^{(3)}$, influences the event 
rates in detectors like Super-Kamiokande, SNO or Borexino.

For oscillations between two neutrino flavors where no mixing into
the third flavor occurs (e.g.~$\nu_e\leftrightarrow\nu_\mu$,
$\Theta_{13}=0$), \mbox{Eq.~(\ref{Pvac})} simplifies to the well known formula 
\begin{equation} \label{Pvac2}
P(r)_{e\rightarrow e}^{(2)} = 1 - \sin^2\left(\!\frac{\Delta m^2 \,r}{4E}\!\right)
\sin^2\!2\Theta ,
\end{equation}
where $\Delta m^2=\Delta m^2_{12}$ or $\Delta m^2_{13}$ for
$\nu_e$-$\nu_\mu$-- resp. $\nu_e$-$\nu_\tau$--oscillations ($\Theta$
defined analogously).

\subsection{Matter effect}

During the propagation of the neutrinos through the Sun they 
coherently scatter forward on the particles of the solar plasma. 
Unlike $\mu$- and $\tau$-neutrinos, which only interact via
NC-reactions, electron-neutrinos can additionally couple via  ${\cal W}$-bosons
to the electrons. Thus, the scattering cross section of a $\nu_e$ is
altered as against the one of the other two 
neutrino-flavors.
This can lead to a resonant flavor 
transition, which may create a pure $\nu_\mu$- or $\nu_\tau$-beam from
the originally 
created electron-neutrinos, first theoretically postulated and
described in \cite{MS} ({\it MSW-effect}).

This effect can be included in \mbox{Eq.~(\ref{schroe})} by substituting ${\cal H}$ with
 $\tilde{\cal  H}$\,=\,${\cal H}+V_{\rm eff}$, where 
\[ 
V_{\rm eff} = \sqrt{2}\,G_F\,N_e\,{\cal U}^\dagger\left(\begin{array}{ccc}
1 & 0 & 0\\
0 & 0 & 0\\
0 & 0 & 0
\end{array}\right){\cal U}
\] 
influences solely the 
$\nu_e$-contribution of the neutrino beam. $G_F$ 
is the Fermi coupling-constant and $N_e$ the electron number density.
The new Hamiltonian $\tilde{\cal H}$ is no longer diagonal in the mass
basis. To evaluate the survival probability and thus
$\exp(-i\tilde{\cal H}t)$ it is therefore necessary to diagonalize
$\tilde{\cal H}$ by a unitary transformation ${\cal V}$ 
\[\tilde{\cal H}_m = \left(\begin{array}{ccc}
M_1 & 0 & 0\\
0 & M_2 & 0 \\
0 & 0 & M_3 
\end{array}\right) = {\cal V}^\dagger \tilde{\cal H} {\cal V}.
\]
Similar to the vacuum
oscillations the constant phase  $M_1\,I\!d_3$ can be removed from
$\tilde{\cal H}$ as it
does not change
the survival probability. The complicated analytical expressions
for $\Delta M_{ij}$ have been evaluated in \cite{Bar80}. 
Recently, $\exp(-i\tilde{\cal H}t)$ has been calculated
in \cite{Ohl00}
by using Cayley-Hamilton's theorem without explicitly deriving
${\cal V}$. In the present work, however, ${\cal V}$ and $M_i$ 
are computed with a fast numerical algorithm using none of the analytic
expressions.

\subsection{The parameter space} \label{pspace}

Recently various publications came up about the actual
size of the necessary parameter space covering all possible solutions
of the solar neutrino problem\cite{Lun99,Fri00,Gou00}. In this section
the sometimes confusing statements
about this topic are summarized and clarified.

The mixing angles can be defined to lie in the first quadrant by
appropriately adjusting the neutrino field phases similar as in the
quark sector~\cite{Cas98}. 
This can also be verified from  
the final formulae, as, for instance, $P(r)^{(3)}_{e\rightarrow e}$ depends
solely on the square of $\sin\Theta_{ij}$ resp.~$\cos\Theta_{ij}$
($ij=12,13$). Moreover, it has been shown in~\cite{Lun99}
that is is also sufficient to consider $0\le\Theta_{23}<\pi/2$, if
$P(r)^{(3)}_{e\rightarrow \mu}$ resp.~$P(r)^{(3)}_{e\rightarrow \tau}$ are to be measured.
In the case of the matter-enhanced
oscillations the situation is not as trivial, but using an analytical
formula for the evolution of a neutrino state in matter as derived in
\cite{Ohl00} (Eq.~44 therein), one can also show that the evolution of
an initial electron-neutrino is determined only by the squares of
$\sin\Theta_{ij}$ and $\cos\Theta_{ij}$.

\subsubsection{Two flavors}
First, the case of
2-flavor oscillations is examined. Exchanging the first and second
row in the definition of the mass eigenstates (Eq.~\ref{eigendef})
implies that $\Delta m^2\rightarrow -\Delta m^2$ and $\sin\Theta \leftrightarrow
\cos\Theta$ ($\Leftrightarrow\Theta\rightarrow(\pi/2 - \Theta)$)\footnote{The
indices 12 or 13 of the mass-squared difference resp.~mixing angle are
omitted in the 2-flavor case.}. Since the 
assignment of the masses $m_i$ to the respective mass eigenstate
$\nu_j$ must not change the results, e.g.~the case $\Delta m^2>0$,
$0\le\Theta<\pi/4$ is equivalent to $\Delta m^2<0$,
$\pi/4\le\Theta<\pi/2$ for any possible form of the Hamiltonian. 
Thus, without loss of generality $\Theta$ can be chosen to be within
$[0,\pi/4]$. 

For pure vacuum oscillations
\mbox{Eq.~(\ref{Pvac2})} can be applied which yields that $P(r)^{(2)}_{e\rightarrow e}$
is independent of the sign of $\Delta m^2$ \emph{and} under the
transformation \mbox{$\Theta\rightarrow(\pi/2-\Theta)$}. Hence, in this case
it is even sufficient to consider \mbox{$\Delta m^2>0$}.

The situation is different for oscillations in matter, where the resonance
condition for MSW-transition is given by
\begin{equation} \label{reson}
N_{\rm res} = \frac{\Delta m^2/2E}{\sqrt{2} G_{\rm F}}\cos\!2\Theta.
\end{equation}
A resonance may occur, if $N_{\rm res}$ is positive, thus 
$\Delta m^2>0$ ($\Theta\le\pi/4$). Although no resonance
occurs for negative values of $\Delta m^2$, the matter still
influences the evolution 
of the neutrino flavor composition in the solar interior~(for a more
detailed study see e.g.~\cite{Bar80}). 

\begin{figure}[t]
\hbox to\hsize{\hss\epsfxsize=8.6cm\epsfbox{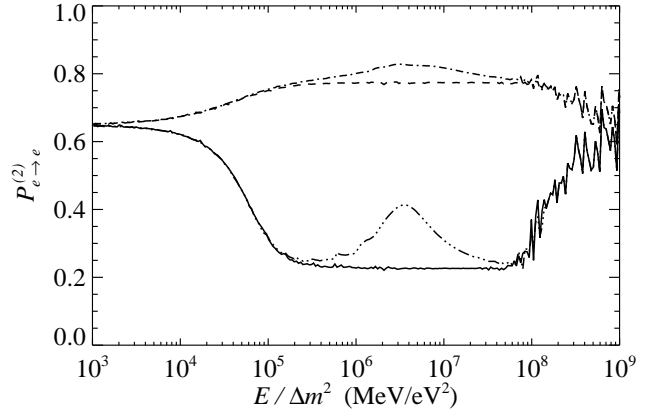}\hss}
\caption{Annually averaged probability of detecting a neutrino
produced in the solar center as
$\nu_e$ on Earth for 
$\sin^22\Theta=0.7$ ($0\le\Theta<\pi/4$) with $\Delta m^2>0$ (solid and
dash-dot-dot-dotted) resp.~$\Delta m^2<0$ (dashed and dash-dotted
line). The two lines in each case show the probability neglecting
(solid and dashed)
resp.~including (dash-dot-dot-dotted and dash-dotted) the Earth regeneration
effect~\protect\cite{Bou86}. \label{bathtub}} 
\end{figure}

In Fig.~\ref{bathtub} the effect of solar matter on the energy 
dependence of $P_{e\rightarrow e}^{\rm (2)}$ is demonstrated.
 For $E/|\Delta m^2|\lesssim 10^8\,{\rm MeV/eV^2}$ vacuum oscillations would yield an
energy-independent survival 
probability for electron-neutrinos, as the vacuum oscillation length
is small and therefore only a mean value of
(Eq.~\ref{Pvac2}) 
\begin{equation} \label{meanvac}
\langle P^{(2)}_{e\rightarrow e}\rangle_{r} = 1-0.5\sin^2\!2\Theta
\end{equation}
would be measurable on Earth. For $\sin^2\!2\Theta=0.7$ this leads to 
$\langle P_{e\rightarrow e}^{\rm (2)}\rangle_{r}=0.65$.
In contrast to the case with \mbox{$\Delta
m^2>0$} (solid line), where the resonant flavor
conversion diminishes the $\nu_e$-contribution of the solar neutrino
current, for $\Delta m^2<0$ (dashed line) the $\nu_e$-portion is even
enhanced compared
to the pure vacuum-oscillation case. Thus, the minimum value of the
$\nu_e$-contribution for $\Delta m^2<0$ is obtained for pure vacuum
oscillations. However, since vacuum oscillations in this mass range
yield at most a suppression of the $\nu_e$-flux of 50\%
(Eq.~\ref{meanvac}) and the solution
to the solar neutrino problem demands, at least to explain the
Homestake experiment, a stronger suppression of about 60\%, for
$\Delta m^2<0$ no reasonable solution can be obtained. 
Hence it is sufficient to consider in the 2-flavor MSW case similar to the
2-flavor vacuum oscillations solely \mbox{$\Delta m^2>0$}
($\Theta\le\pi/4$). 

Recently, it has been pointed out in \cite{Gou00,Gon00}
that the $3\sigma$-ranges of the
LMA- and LOW-solutions (see below) extend outside this region,
but additional solutions for the solar neutrino
problems cannot be found there. In their analysis, however, 
$\Delta m^2$ was fixed to be 
positive and thus $0\le\Theta<\pi/2$ has been examined. The region
$\pi/4<\Theta<\pi/2$ of their parameter space (termed ``the dark side'')
is equivalent to the region $\Delta m^2<0$ and $0\le\Theta<\pi/4$
discussed above.  

\subsubsection{Three flavors}
\begin{figure}[ht]
\hbox to\hsize{\hss\epsfxsize=8.6cm\epsfbox{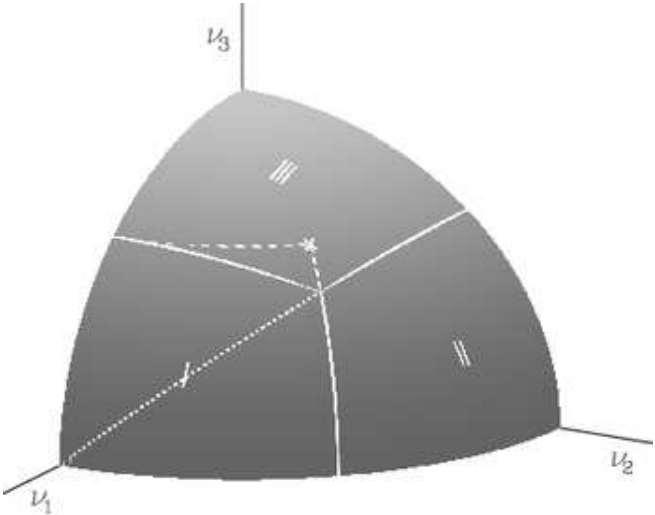}\hss}
\caption{Geometrical illustration of the flavor space available for a
$\nu_e$ in the mass basis $\left(\nu_1,
\nu_2, \nu_3\right)^{\rm T}$. The spherical segment is divided into three
sectors of equal size and shape, where each can be obtained from
another by mirroring at the respective diagonal. Maximal mixing is
given at the intersection of all 
three sector borders, i.e.~$\sin^2\!2\Theta_{12}=1$ and
$\sin^2\!2\Theta_{13}=\frac{8}{9}$. The cross in \mbox{\rm sector {\small III}} indicates
the case $\sin^2\!2\Theta_{12}=\sin^2\!2\Theta_{13}=1$, the dashed 
border extends \mbox{\rm sector {\small I}} to the area with $\Theta_{12}, \Theta_{13} \le
\pi/4$. The dotted line splits \mbox{\rm sector {\small I}} into two symmetric parts. (See
text for more details.) 
\label{spher}}
\end{figure}

The considerations of the 2-flavor case are now extended
to three flavors. In Fig.~\ref{spher} the flavor space for the
electron-neutrino is illustrated, where $\nu_e$ is represented as a
(yet unknown) point on the surface of an eight of an unit sphere. 
While in the 2-flavor case the ordering of the masses enables two
cases to be distinguished ($\Delta m^2<0$ resp.~$\Delta m^2>0$), in the
3-flavor scenario six cases can be identified. But, since
exchanging any axes in Fig.~\ref{spher} maps the flavor space 
onto itself, each mass hierarchy can of course be obtained from the
``canonical'' 
one ($m^2_1< m^2_2< m^2_3$) simply by exchanging the respective
assignment of $m_i$ to $\nu_j$ in \mbox{Eq.~(\ref{eigendef})}.

Unlike the 2-flavor case, where for pure vacuum
oscillations the parameter space could be further decreased, it is not
possible in the general 3-flavor scenario, as e.g.~$\Delta m^2_{23} =
\Delta m^2_{13}-\Delta m^2_{12}$, is not invariant under the
transformation $\Delta m^2_{12}\rightarrow-\Delta m^2_{12}$.

For 3-flavor neutrino oscillations in matter an exact analytic
resonance condition can hardly be obtained because of the complicated
formula for the mass eigenvalues $\Delta M_{ij}$.
If the masses are well separated, the 2-flavor resonance condition
(Eq.~\ref{reson}) can be applied for both systems by substituting the
quantities ($\Delta m^2, \cos\!2\Theta$) by ($\Delta m^2_{12}\,
\cos\Theta_{13}, \cos\!2\Theta_{12}$)   
resp.~($\Delta m^2_{13}, \cos\!2\Theta_{13}$)~\cite{Kuo86}. Hence, in 
this case considering $\Theta_{12},\Theta_{13}\le\pi/4$ (\mbox{\rm
sector {\small I}})
extended to the dashed borders in \mbox{\rm sector {\small III}} of Fig.~\ref{spher}) with
$\Delta m^2_{12},\Delta m^2_{13}\ge0$ is sufficient to obtain all 
possible solutions where two resonances may occur. However, the
solution to the solar neutrino problem may also 
be a combination of a non-resonant and a resonant oscillation and the
masses can be in principle very similar, too. Thus, the whole
parameter space must be taken into account.

In \cite{Fog96}, for instance, 3-flavor oscillations were investigated
assuming canonical mass hierarchy and 
using a fixed value for $\Delta m^2_{13}=10^{-3}~{\rm eV^2}$ in agreement
with the atmospheric-neutrino results. They examined the remaining
parameter space applying an analytical formula for the
$\nu_e$ survival probability, which approximates the solar electron-density 
profile by an exponential function and is valid for  
$\Delta m^2_{13}\approx 10^{-3}~{\rm eV^2}$ and $\Delta m^2_{12}$ significantly
smaller than $\Delta m^2_{13}$. Under these conditions the parameter space
could be reduced considerably. 

In the present work, the most general case of 3-flavor oscillations of
solar neutrinos is investigated, and thus these restrictions are not
applicable here. Instead of performing a survey over the
whole flavor space ($0\le\Theta_{12}, \Theta_{13}<\pi/2$) with
canonical mass hierarchy, I prefer to consider 
the case $\Delta m^2_{12}, \Delta m^2_{13}>0$ which covers two possible
mass hierarchies. Thus, in this case only half of the total flavor space
must be overviewed. In the illustration provided in Fig.~\ref{spher}
this reduced area is given by \mbox{\rm sector {\small II}} and the lower half of 
\mbox{\rm sector {\small I}} defined by the dotted line.

To simplify the numerical survey and the analysis a variable
transformation to the 
mixing angles is applied in each sector:
Obviously $P_{e\rightarrow e}^{(3)}$ as defined in \mbox{Eq.~(\ref{Pvac})} is not
symmetric under the exchange 
of the indices $2\leftrightarrow 3$. However, \mbox{\rm sector {\small I}} is symmetric under
$\nu_2\!\leftrightarrow\!\nu_3$, and it would be useful to  
have two quantities $\tilde{\Theta}_{12}(\Theta_{12},\Theta_{13})$ and
$\tilde{\Theta}_{13}(\Theta_{12},\Theta_{13})$ which fulfill
\begin{eqnarray*}
P^{(3)}_{e\rightarrow e}(\Delta m^2_{12},\tilde{\Theta}_{12},\Delta
m^2_{13},\tilde{\Theta}_{13}) & = &\\ 
&&\hspace{-1.5cm}P^{(3)}_{e\rightarrow
e}(\Delta m^2_{13},\tilde{\Theta}_{13},\Delta
m^2_{12},\tilde{\Theta}_{12}). 
\end{eqnarray*}
Defining in \mbox{\rm sector {\small I}} ($0\le\Theta_{12}<\pi/4$,
$\tan\Theta_{13}\le\cos\Theta_{12}$) the quantities
$\tilde{s}_{12}$(=\,$\sin\tilde{\Theta}_{12}$)  and $\tilde{s}_{13}$
by
\begin{equation}\label{angsym}
 \begin{minipage}{5cm}
\begin{eqnarray*}
\tilde{s}_{12} & = & s_{12} \\
\tilde{s}_{13} & = & \frac{s_{13}}{\sqrt{1-s_{12}^2c_{13}^2}},
\end{eqnarray*}\end{minipage}
\end{equation}
yields such a pair. With these quantities \mbox{Eq.~(\ref{Pvac})} writes as ($\tilde{c}_{ij}^2=1-\tilde{s}_{ij}^2$)
\begin{eqnarray*} 
P(r)_{e\rightarrow e}^{(3)}  & = & 1 - \left(\frac{2\tilde{c}_{12}\tilde{c}_{13}}  
{1\;-\tilde{s}^2_{12}\tilde{s}^2_{13}}\right)^2
 \Biggl[\tilde{s}^2_{12}\tilde{c}^2_{13}\, \sin^2\left(\!\frac{\Delta
m^2_{12}\,r}{4E}\!\right) \\
&&\hspace*{-1cm}+  \tilde{c}^2_{12}\tilde{s}^2_{13}\, \sin^2\left(\!\frac{\Delta m^2_{13}\,r}{4E}\!\right)
+  \tilde{s}^2_{12}\tilde{s}^2_{13}\, \sin^2\left(\!\frac{\Delta m^2_{23}\,r}{4E}\!\right)\Biggr],
\end{eqnarray*}
which is unaltered under the exchange of the indices \mbox{$2\leftrightarrow3$}.

Moreover, it can be proven that the evolution matrix of neutrinos in
matter is not changed when exchanging the indices
$2\leftrightarrow3$\cite{PhD}. Hence, using $\tilde{s}_{12}$  
and $\tilde{s}_{13}$ in \mbox{\rm sector {\small I}} yields a $\nu_e$ survival probability,
which is constant under this transformation in vacuum as well as in
matter. 

Similar to \mbox{\rm sector {\small I}} also \mbox{\rm sector {\small II}} ($\pi/4\le\Theta_{12}<\pi/2$,
$\tan\Theta_{13}\le\sin\Theta_{12}$) can be 
brought in a quadratic shape by using the transformation 
\begin{equation} \label{angII}
\tilde{s}_{13} = \frac{s_{13}}{\sqrt{1-c_{12}^2c_{13}^2}}.
\end{equation}
But this does not yield a survival probability
invariant under the exchange of the indices \mbox{$2\leftrightarrow3$}, as under
this operation \mbox{\rm sector {\small II}} would be mapped on \mbox{\rm sector {\small III}}.

\section{Calculations}\label{calc}

\subsection{Standard solar model}
For the following calculations my standard solar model (``GARSOM4'')
as described in~\cite{SW99} is used. It has been calculated
using the latest input physics, 
equation of state and opacity from the OPAL-group \cite{OPEOS,OP96}, and
nuclear reaction rates as proposed by~\cite{Adel98}. In addition, 
microscopic diffusion of H, $^3$He, $^4$He, the CNO-isotopes, and 4
heavier elements (among them Fe)
is included using the diffusion constants of \cite{Diffc}. By treating
convection as a fast diffusive process the chemical changes due to
nuclear burning and diffusion (mixing) are evaluated in a common
scheme. 

A peculiarity of GARSOM4 is the inclusion of realistic
2d-hydrodynamical model atmospheres~\cite{2dhydro} until an optical
depth of 1000. The improvement of the high-degree p-mode frequencies
due to a better reproduction of the superadiabatic layers
just below the photosphere is similar to the one obtained by using
1d-model atmospheres like in \cite{SWL97}.  The advantage of using the
2d- instead of the 1d-model atmospheres is the extension of the former to
greater optical depths, where the stratification is already adiabatic
and thus the solar model gets nearly independent of the applied convection
theory. 

\begin{table}[b]
\squeezetable
\caption{Typical quantities of GARSOM4. $\alpha_{\rm CM}$ is the
mixing-length parameter of the convection theory developed
in~\protect\cite{CM}. $Y$ and $Z$ are helium and metal mass
fractions, $T$ and $\rho$ temperature and mass density, respectively.
The  depth of the 
convective envelope in units of the solar radius is abbreviated by 
$R_{\rm  bce}$. The indices $i$, $s$, and $c$ denote initial,
surface (photospheric) and central values, respectively.\label{smodel}} 
\begin{tabular}{ccccccccc}
&&&&& $T_c$ & $\rho_c$ & & $\rho_{\rm bce}$ \\ 
\rbox{$\alpha_{\rm CM}$} & \rbox{$Y_i$} & \rbox{$Z_i$} & \rbox{$Y_s$}
&\rbox{$Z_s$} & $(10^7\,{\rm K})$ & ${\rm (g/cm^3)}$ &
\rbox{$R_{\rm bce}$} & ${\rm (g/cm^3)}$ \\ \hline
0.975 & 0.275  & 0.020  & 0.245  & 0.018  &1.57 & 152 & 0.713  & 0.188  
\end{tabular}
\end{table}

In Table~\ref{smodel} typical quantities of GARSOM4 using the
convection theory developed in~\cite{CM} are summarized.
The predicted event rates of GARSOM4 for the three
presently operating types of neutrino experiments GALLEX/GNO/SAGE,
Homestake, and Super-Kamiokande are summarized in Table~\ref{neurate} together
with the values obtained by~\cite{Bah98}. Since the input physics is very
similar in both models, the predicted rates agree very well within the
errors as quoted by~\cite{Bah98}. In addition, the influence of the
assumed $^8$B-neutrino spectrum to the predicted rates is shown. Using
the recently measured spectrum~\cite{OGW00} yields slightly higher rates for
all three experiments than including the theoretically
predicted one~\cite{Bah96}, which is caused by the somewhat larger number of high-energetic
$^8$B-neutrinos in the former spectrum and 
the strongly inclining detection probability toward higher energies.

\begin{figure}[t]
\hbox to\hsize{\hss\epsfxsize=8.6cm\epsfbox{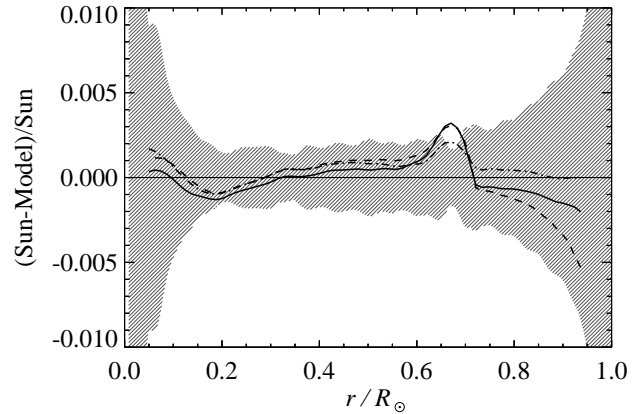}\hss}
\caption{Comparison of modern solar models with the seismic model by
\protect\cite{Bas98}: shown is the relative 
(seismic-solar model) difference of sound speed for the model of reference
\protect\cite{Bah98} (dashed line), reference \protect\cite{sci0696}
(dash-dotted), and GARSOM4 (solid). The grey-shaded
area indicates a conservative error range of seismic models according to
\protect\cite{Sci97}.\label{udiff}}
\end{figure}

Figure~\ref{udiff} shows the sound-speed profile of GARSOM4 compared to the
seismic model inferred by \cite{Bas98}. The deviations of
GARSOM4 from the latter are of the same size as standard solar models
from other groups~\cite{Bah98,sci0696}. 

\begin{table}[b]
\caption{Expected event rates
resp.~neutrino flux in the three types of experiments 
GALLEX/GNO/SAGE (Ga), Homestake (Cl), and Super-Kamiokande as predicted by the two
solar models GARSOM4 and BP98~\protect\cite{Bah98}. For GARSOM4 the
expected values using the measured~\protect\cite{OGW00} (first row)
and the theoretically derived~\protect\cite{Bah96} (second
row) $^8$B-neutrino spectrum are given. The last row provides the 
the measured values with their respective errors.\label{neurate}}
\renewcommand{\arraystretch}{1.2}
\begin{tabular}{cccc}
  & Ga [SNU] & Cl [SNU] & Super-K [$10^6 {\rm cm^{-2}s^{-1}}$]
\\ \hline
               & 128.7 & 7.79 &  5.18 \\
\rbox{GARSOM4} & 128.4 & 7.58 &  5.06 \\ \hline
BP98    & $129^{+8}_{-6}$ & $7.7^{+1.2}_{-1.0}$ &
$5.15^{+1.00}_{-0.72}$ \\ 
Experiment  &
$74.2\pm4.9$\tablenote{Reference~\cite{GNO00}} & 
$2.56\pm0.23$\tablenote{Reference~\cite{Home98}} &
$2.40\pm0.08$\tablenote{Reference~\cite{Suz00}} \\
\end{tabular}
\end{table}

\subsection{Neutrino-oscillation analysis}

Using the neutrino flux as provided by GARSOM4 the evolution of the
initial electron neutrinos through Sun, space and Earth is
computed taking into account oscillations between the flavors (see
Appendix). The electron-density profile and the radial
distribution of the neutrinos is taken from the solar model, too.
For the electron-density profile of the Earth the
spherically symmetric PREM-model~\cite{PREM} is applied.
For each set of mixing parameters the neutrino energy spectrum
observed on Earth is evaluated and folded with the detector
response functions. The combinations of mixing parameters which reproduce the
measured data are found by applying a $\chi^2$-analysis. 

There are
four contributions to the total value of $\chi^2$ originating from the
four different available data sets, the event rates, the
recoil-electron energy spectrum, the zenith-angle distribution, and the
annual variation. The latter three are available only from the Super-Kamiokande
detector, while to the first one all three types of neutrino
experiments contribute.

For the event-rate portion the commonly used formula holds 
\begin{equation}\label{chir}
\chi^2_R = \sum_{i=1}^{3}\frac{\left(N_i^{\rm exp}-N_i^{\rm
th}\right)^2}{\sigma_{i,{\rm exp}}^2 + \sigma_{i,{\rm th}}^2},
\end{equation}
where $i$ denotes GALLEX/GNO/SAGE, Homestake or Super-Kamiokande. $\sigma_{\rm
exp}^2$ and $\sigma_{\rm th}^2$ are the 
experimental resp.~theoretical 1$\sigma$-errors
(Table~\ref{neurate}). Since the input physics in 
GARSOM4 is similar to the one used in \cite{Bah98}, the
theoretical errors derived therein are taken for $\sigma_{\rm th}^2$. 
$N_i^{\rm exp}$ are the measured event rates, which are quoted
together with the uncertainties in Table \ref{neurate}.  Note, that the
Super-Kamiokande 
data are usually reported as $^8$B-neutrino flux relatively to a
standard solar-model prediction. Actually this number has to be
understood as an event-rate ratio. The total number of measured events
are divided by the theoretically expected value (e.g.~from GARSOM4). This
ratio is then often  falsely taken to be the suppression rate of the
\emph{total} 
$^8$B-neutrino flux. However, with the 
energy window of the recoiled electrons being between
5.5\footnote{There are already data for the energy window from 5.0 to
5.5\,MeV available, but the systematic errors are still to
be derived.}  
and 20\,MeV no statement about the total number of $^8$B-neutrinos
below this window is possible. Moreover, neutrino oscillations may
alter the energy spectrum of the $^8$B electron-neutrino flux and as the
scattering cross section of the neutrinos in Super-Kamiokande is
energy-dependent, the same number of event rates can be obtained with 
different $^8$B-neutrino fluxes. Thus in the present analysis
the event rate in Super-Kamiokande following \mbox{Eq.~(\ref{skrate})} are used
and not the total $^8$B-neutrino flux.

The recoil-electron energy spectrum is examined by 
\begin{equation} \label{chie}
\chi^2_e = \sum_{i=1}^{18} \left(\frac{N_{i,e}^{\rm exp} -
\alpha_e N_{i,{\rm e}}^{\rm th}}{\sigma_{i,e}}\right)^2,
\end{equation}
where the sum extends over all 18 energy bins (Fig.~\ref{esp_msw2}(a))
and $\sigma_{i,e}$ is the quadratic sum of statistical and
systematic errors taken from~\cite{Suz00}.
Since the absolute value of the event rate in Super-Kamiokande has already been used in 
$\chi^2_R$ the parameter $\alpha_e$ is introduced, by which
the spectrum can be normalized adequately, independent of the total rates.

The contribution of the zenith-angle dependence (6 bins,
Fig.~\ref{esp_msw2}(b)) and seasonal-variation data (4 bins,
Fig.~\ref{esp_msw2}(c)) is defined analogously 
(denoted in the following $\chi^2_Z$ resp.~$\chi^2_A$).

\begin{figure}[b]
\hbox to\hsize{\hss\epsfxsize=8.6cm\epsfbox{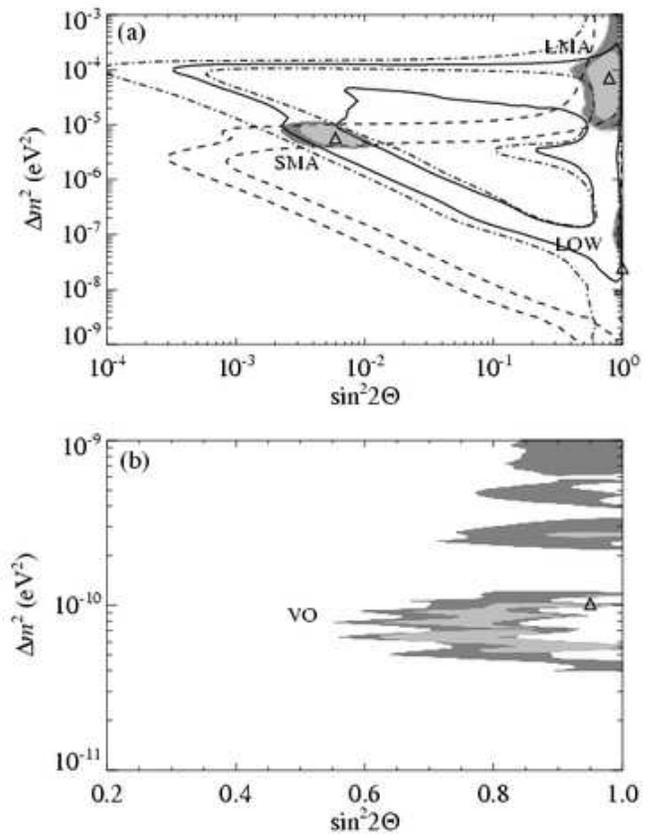}\hss}
\caption{Confidence regions in the $\Delta m^2$-$\sin^2\!2\Theta$--plane
for 2-flavor neutrino oscillations using only the event rates. Due
to the different mass ranges and dependences on the mixing angle, the
MSW-solutions are shown in (a), the pure
vacuum-oscillation solutions in (b). The 
light shaded areas reflect the  
95.4\% C.L., while the dark shaded regions show the 99.7\% C.L. Also 
drawn in (a) are the constraints from Homestake (solid),
GALLEX/GNO/SAGE (dashed),
and Super-Kamiokande (dash-dotted line). The best-fit value
of each solution is indicated by $\triangle$ (Table
\ref{bestmsw2}).\label{msw2}} 
\end{figure}

\section{Results}\label{results}

\subsection{Two flavors}\label{twoflav}

In a first step the solar neutrino problem is analyzed taking into
account oscillations only between two flavors. Figure \ref{msw2} shows
the allowed oscillation parameters using the experimentally derived
$^8$B-neutrino spectrum~\cite{OGW00}, if solely the event rates of the 
three experiments are fitted. Clearly the four commonly known
solution-islands~(see e.g.~in \cite{Hat97,BahK97}) can be
identified, the small-mixing (SMA) and large-mixing angle (LMA), the
low mass-squared difference (LOW), and the vacuum-oscillation (VO) solutions. 
The $\chi^2$-values for the best-fit parameters in these solutions are
quoted in Table~\ref{bestmsw2}. 
Apart from LOW all solutions have a ratio of
$\chi^2$ to the number of degrees-of-freedom (d.o.f.) which is less than
one and therefore these solutions are acceptable candidates as 
correct solution for the solar neutrino puzzle.

In the last four columns in Table~\ref{bestmsw2} the best-fit values
using the theoretically derived $^8$B-neutrino spectrum of
reference\cite{Bah96} 
are provided. The somewhat higher expected event rates 
(Table \ref{neurate}) result in slightly different mixing
parameters compared to the case using the 
measured $^8$B-neutrino spectrum~\cite{OGW00} (first four 
rows). While the $\chi^2$-values for the SMA-, LMA-, and VO-solutions 
are marginally worse with the theoretical spectrum, the LOW-solution
give a slightly better fit to the experiments. However the changes are in
all cases relatively small.  

\begin{table}[t]
\caption{Best-fit values for oscillations between two neutrinos taking
into account solely the event rates of the three detector types.  In
the first four rows the measured $^8$B-neutrino
spectrum~\protect\cite{OGW00} is included, while in the last rows the
spectrum calculated in~\protect\cite{Bah96} is assumed.
The total number
of d.o.f in this analysis is 1.\label{bestmsw2}}  
\renewcommand{\arraystretch}{1.1}
\begin{tabular}{clcd}
& \multicolumn{1}{c}{$\Delta m^2$ (eV$^2$)} &
\multicolumn{1}{c}{$\sin^2\!2\Theta$} &
\multicolumn{1}{c}{$\chi^2_{\rm tot}$} \\ \hline
SMA & $5.7\times10^{-6}$ & $5.9\times10^{-3}$ & 0.16 \\
LMA & $7.1\times10^{-5}$ & 0.79 & 0.79 \\
LOW & $2.5\times10^{-8}$ & 1.00 & 5.00 \\
VO  & $1.0\times10^{-10}$ & 0.95 & 0.16 \\ \hline
SMA & $5.4\times10^{-6}$ & $6.1\times10^{-3}$ & 0.21 \\
LMA & $7.3\times10^{-5}$ & 0.78 & 0.81 \\
LOW & $5.0\times10^{-8}$ & 1.00 & 3.70 \\
VO  & $8.1\times10^{-11}$ & 0.73 & 0.61 \\ 
\end{tabular}
\end{table}

While the event-rates alone favor the SMA-solution, including 
into the analysis the recoil-electron energy spectrum, the
zenith-angle and annual variations recorded by Super-Kamiokande,
yields the LMA-solution as the best fit (Table~\ref{totmsw2}). This
result could also be found with the previous 825-day Super-Kamiokande
data (see e.g.~in\cite{Bah99}). In contrast to the earlier analyses, where no
set of parameter could be found, which reproduces all sets of
data at the same time~\cite{PhD}, with the 1117-days data such
simultaneous fits can be performed. In Table~\ref{totmsw2} the
best-fit values of these solutions, the respective $\chi^2$-contributions
and the ratio of $\chi^2_{\rm tot}$ to the available d.o.f. are
provided.

\begin{figure}[ht]
\hbox to\hsize{\hss\epsfxsize=8.6cm\epsfbox{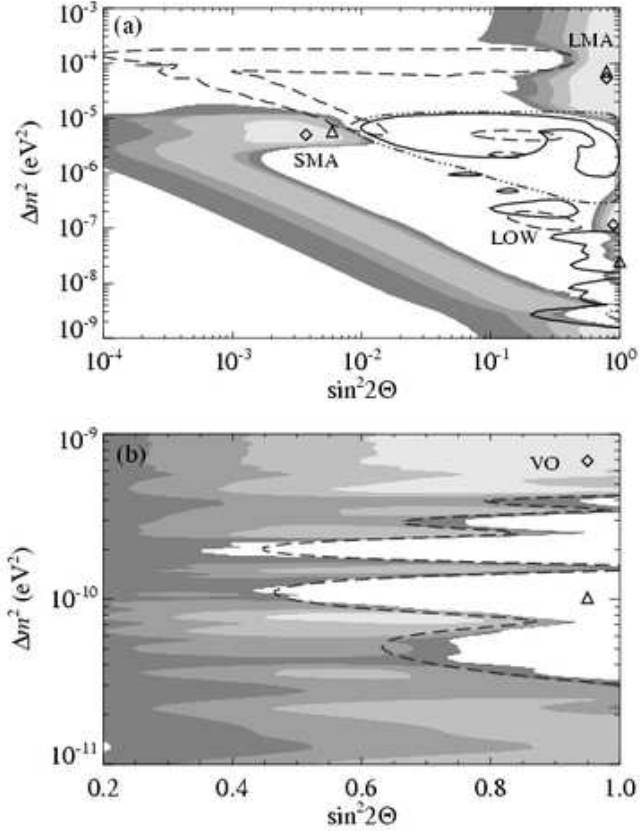}\hss}
\caption{Confidence regions in the $\Delta m^2$-$\sin^2\!2\Theta$--plane
for 2-flavor neutrino oscillations using all data. The 
different shaded areas reflect from light to dark, 10, 63.7, 
95.4, and 99.7\% C.L. Also 
drawn in (a) are the constraints from the recoil-electron energy
spectrum (dashed), 
zenith-angle (dashed-dotted)
and annual variation (solid line) of the Super-Kamiokande data. The best-fit value
of this analysis is indicated by $\diamond$ (Table \ref{totmsw2}); the
$\triangle$ show the best-fit values of the analysis taking only the
event-rate data (Table \ref{bestmsw2}).\label{mswfull2}} 
\end{figure}

\begin{figure}[t]
\hbox to\hsize{\hss\epsfxsize=8.6cm\epsfbox{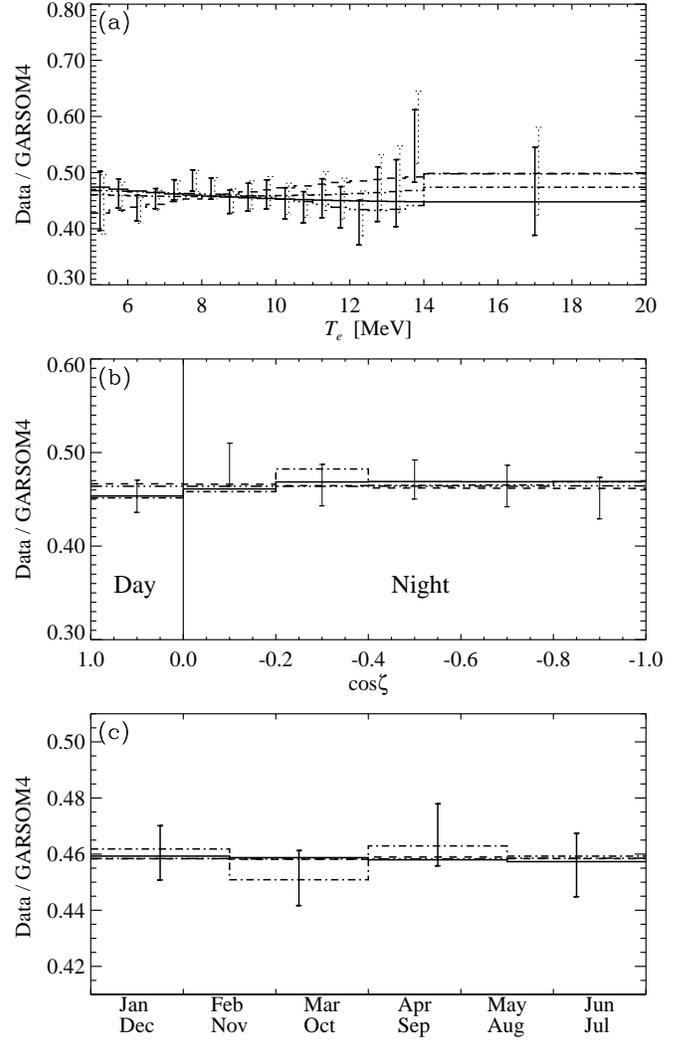}\hss}
\vspace*{2mm}
\caption{Recoil-electron energy spectrum (a), zenith-angle ($\zeta$)
variation (b), and seasonal dependence of the best-fit LMA- (solid line),
SMA- (dashed), LOW- (dash-dotted), and VO- (dash-dot-dot-dotted) solutions
in the 2-flavor oscillation case
including all available solar 
neutrino data (Table~\ref{bestmsw2}, lower part) compared with the 
measurements of \mbox{Super-Kamiokande} (1117-days
data~\protect\cite{Suz00}) using the measured $^8$B-neutrino spectrum
\protect\cite{OGW00} (solid) resp.~the theoretically derived one \protect\cite{Bah96} (dotted bins). All bins are equally spaced in the 
respective data area apart from the last bin in the energy spectrum,
the range of which extends from 14 to 20\,MeV.  The flux variation due
to the eccentric orbit of the Earth has been subtracted in
(c).\label{esp_msw2}}
\end{figure}

The 1, 2, and 3$\sigma$-regions (63.7, 95.4, and 99.7\% C.L.)
including all available data (Fig.~\ref{mswfull2}) are much bigger than
in the case of taking solely the event rates (Fig.~\ref{msw2}). Most
probably neglecting the correlations between the different data sets
of Super-Kamiokande has caused this growth. More detailed data
published from the Super-Kamiokande collaboration are desirable to be
able to perform a more accurate analysis of the solar neutrino data. 
In addition, theoretical correlations between e.g.~the annual and
zenith-angle data should be taken into consideration. In order to
show, that the most probable solution regions have indeed not changed
drastically compared with the pure-rate analysis, the 10\% C.L.~areas
are plotted 
in Fig.~\ref{mswfull2}, too. The same region would result as the
1$\sigma$-area, if the 
minimum $\chi^2_{\rm tot}$-value would be zero instead of 12.5
(=$0.48\times26$, Table~\ref{totmsw2}).

Independent of whether the theoretically derived or measured
$^8$B-neutrino spectrum is used, the LMA-solution can reproduce each single
kind of data, the event rates, the recoil-electron energy spectrum, the
zenith-angle and the seasonal variation acceptably. Although, 
$\chi_{\rm tot}^2/{\rm d.o.f.}$ of the LOW- and VO-solutions are less than
one, these solution do not lead to an acceptable fit of the event-rate
data. Whether the SMA-solution can already be ruled out by the new
Super-Kamiokande data depends on the $^8$B-neutrino spectrum included in the
analysis.
While with the theoretically derived spectrum~\cite{Bah96}, the
event rates can just be reproduced ($\chi^2_{\rm R}/d.o.f.\approx 1$),
including the measured one~\cite{OGW00} hardly yields a
reasonable fit.

\begin{table}[t]
\squeezetable
\caption{Best-fit values for 2-flavor neutrino oscillations and the respective
$\chi^2$-values taking into account all available data sets (event rates of GALLEX/GNO/SAGE, Homestake, and Super-Kamiokande and  zenith-angle
dependence, annual variation, and recoil-electron energy spectrum data
of the latter). 
In the first four rows the measured $^8$B-neutrino spectrum is
utilized~\protect\cite{OGW00}, while in the latter rows the spectrum calculated in~\protect\cite{Bah96} is assumed. $\chi^2_R$, $\chi^2_e$, $\chi^2_Z$ and
$\chi^2_A$ are the individual contributions to $\chi^2_{\rm tot}$ as
defined by Eqs.~(\ref{chir}) and (\ref{chie}).
The last row specifies the contribution of each
$\chi^2$-portion to the total number of d.o.f; the total 
number of d.o.f in this analysis is 26.\label{totmsw2}} 
\begin{tabular}{clcddddd}
& \multicolumn{1}{c}{$\Delta m^2$ (eV$^2$)} &
\multicolumn{1}{c}{$\sin^2\!2\Theta$} &
\multicolumn{1}{c}{$\chi^2_R$} & \multicolumn{1}{c}{$\chi^2_e$}  
& \multicolumn{1}{c}{$\chi^2_Z$} &
\multicolumn{1}{c}{$\chi^2_A$} & \multicolumn{1}{c}{$\chi^2_{\rm
tot}$}/d.o.f. \\ \hline  
SMA & $5.0\times10^{-6}$ & $3.7\times10^{-3}$ & 6.1 & 16.2 & 1.8 & 1.1 & 0.97 \\
LMA & $5.3\times10^{-5}$ & 0.79 & 0.9 & 8.6 & 2.0 & 1.2 & 0.48 \\
LOW & $1.1\times10^{-7}$ & 0.89 & 6.0 & 8.6 & 2.9 & 0.1 & 0.68 \\
VO & $6.9\times10^{-10}$& 0.95 & 6.8 & 8.0 & 1.9 & 1.4 & 0.70 \\ \hline
SMA & $5.4\times10^{-6}$ & $4.2\times10^{-3}$ & 3.3 & 13.5 & 2.1 & 2.0 & 0.79 \\
LMA & $5.0\times10^{-5}$ & 0.88 & 2.0 & 10.9 & 1.8 & 1.7 & 0.63 \\
LOW & $1.0\times10^{-7}$ & 1.00 & 5.9 & 10.1 & 2.7 & 0.7 & 0.75 \\ 
VO & $8.6\times10^{-10}$ & 0.99 & 6.4 &  7.9 & 2.1 & 1.5 & 0.68 \\ \hline
d.o.f. &&& 3 & 17 & 5 & 3 &
\end{tabular}
\end{table}

By assuming that the Chlorine rate is due to unknown systematic errors 
30\% higher than quoted and by reducing the reaction rate for the
$^7$Be-proton capture by about 15--20\%, a VO-solution could be obtained 
which would be able to reproduce the event rates {\it and} the
energy spectrum of Super-Kamiokande fairly well~\cite{Ber00}. However,
presently 
there is no evidence for any hidden systematic uncertainties in
Homestake, which would increase the event rate to the 3.5$\sigma$-level.

The parameters of the best-fit LMA- and SMA-solutions are 
weakly modified when including all the data in the analysis
(Tables~\ref{bestmsw2} and \ref{totmsw2}). The recoil-electron energy spectrum mainly 
influences the SMA-solution, while the zenith-angle data causes a
slight shift of the best-fit values of the LMA-solution
(Fig.~\ref{mswfull2}). 

For $\Delta m^2\approx 10^{-7}$--$10^{-5}$\,eV$^2$ the seasonal
variations are caused by the zenith-angle variation (on the northern
hemisphere more night data are recorded during winter than during
summer), and thus no additional constraints can be obtained from the
former data set in this parameter space. 
For smaller $\Delta m^2$ the eccentricity of the Earth orbit leads to
a ``real'' annual dependent signal, which can be used 
to constrain the mixing parameters. Note, that in contrast to the $\Delta 
m^2\gtrsim10^{-7}$--region the seasonal dependence in  $\Delta
m^2\lesssim10^{-7}$ is now producing tiny day-night variations.

Anyway, for $\Delta
m^2\lesssim10^{-9}\,{\rm eV}^2$ deviations of less than
2\% from an annually constant neutrino
flux are predicted for the Super-Kamiokande data\footnote{In 
the analysis of the seasonal data the neutrino signal has been corrected
for the $1/r^2$-dependence of the flux.}, which is consistent within the
errors with the recorded value.   
Thus, for the VO-solutions only very weak constraints can be obtained
from the present seasonal variation data,
while for the region of the LOW-solution these data provide important 
information. In fact, the position of the best-fit value of the LOW-solution
has been changed by including the annual-variation data in the analysis
(Fig.~\ref{mswfull2}).

In the regime of the VO-solution, the recoil-electron energy spectrum
provides very stringent constraints on the allowed mixing parameters,
excluding great part of the region favored by the rates. Hence, no
good solution in the VO-region could be found, which reproduces the
recoil-electron energy spectrum as well as the rates recorded in
GALLEX/SAGE/GNO, Homestake, and Super-Kamiokande.

Nevertheless, the LMA-solution is presently the favored solution to
the solar neutrino problem, whereas the earlier favored
SMA-solution seem to be almost ruled out. But still, improved statistics in
the recoil-electron data of Super-Kamiokande is required to
identify more reliably the correct solution to the
neutrino problem. 

\subsection{Three flavors}

In spite of the LMA-solution being able to reproduce 
the recoil-electron energy spectrum, the zenith-angle and
seasonal variations, and the event rates acceptably, 
the analysis is extended to all three families to deduce whether a
better fit to the data can be achieved. Besides, this is the physically correct
treatment, which contains the 2-flavor case as a limiting one.

The electron-neutrino survival probability for 3-flavor
neutrino oscillations is determined by four  
quantities $\Delta m^2_{12}$, $\sin^22\Theta_{12}$, $\Delta
m^2_{13}$, and $\sin^22\Theta_{13}$, where the appropriate 
pairs describe each the mixing of two flavors. Hence a 4-dimensional
parameter space has to be examined to deduce all possible solutions. In the
Appendix the numerical realization is described with
which the 4-dimensional parameter survey can be performed
efficiently. As worked out in section \ref{pspace}, all possible solutions
for MSW-solutions are obtained by considering \mbox{\rm sector {\small I}} and {\small II}
(Fig.~\ref{spher}) with $\Delta m^2_{12}$, $\Delta m^2_{13}>0$. Using
in \mbox{\rm sector {\small I}} the quantity
$\sin^22\tilde{\Theta}_{13}$ as defined
in \mbox{Eq.~(\ref{angsym})} instead of $\sin^22\Theta_{13}$, allows
to describe the pure 2-neutrino 
$\nu_1$-$\nu_3$--oscillations equivalently to the
$\nu_1$-$\nu_2$--case, which has been investigated thoroughly in various
publications~\cite{Hat97,BahK97}. 
In addition, with the survival probability
$P(r)_{\rm e\rightarrow e}^{(3)}(\Delta m^2_{12}, \sin^22\Theta_{12}, \Delta
m^2_{13}, \sin^22\tilde{\Theta}_{13})$ being symmetric in 
the exchange of the indices 2 and 3 unnecessary computations can be
avoided. Nevertheless, the computations are very extensive due to the
4-dimensional parameter space. Therefore, the grid in the 3-flavor
oscillation-survey has to be chosen less dense than 
in the 2-flavor case, where only a 2-dimensional grid
had to be overviewed. However, the grid must still be fine
enough that those solutions are not missed which might be confined to 
small regions in the parameter space.
Furthermore, the number of neutrino paths from the Sun to
the detector to cover the whole year of data recording was
reduced compared to the pure 2-flavor neutrino oscillations. This 
leads to slightly different $\chi^2$-values for effectively pure
2-flavor solutions, which are also found in the full analysis. 

Subspaces of the entire possible parameter space have been
investigated e.g.~in \cite{Fog96} or \cite{Liu97}. In both
publications an analytical expression for the survival probability
$P_{e\rightarrow e}^{(3)}$ derived in \cite{Pet88} has been used, which is valid for
large mass separations resp.~small mixing angles. In this description
$P_{e\rightarrow e}^{(3)}$ is determined by two 2-flavor probabilities
$P_{e\rightarrow e}^{(2)}$ for each mass splitting $\Delta m^2_{12}$  and
$\Delta m^2_{13}$.  
By approximating the electron-density profile in the solar interior
with an exponential function, $P_{e\rightarrow e}^{(2)}$ can also be evaluated
analytically~\cite{Pet88}.

The results obtained in \cite{Fog96} or \cite{Liu97}
could always been reproduced in the respective mass ranges. However, 
in those investigations of the 3-flavor scenario, only the event-rate
data were 
available. Using the new type of data (recoil-electron energy spectrum,
zenith-angle and seasonal variations) almost all solutions found
in~\cite{Fog96,Liu97} are disfavored. Furthermore, new solutions are 
identified since the respective parts of the parameter space were not
covered in the analyses therein.

In the $\chi^2$-analysis, applied to constrain the mixing parameters,
always the whole available experimental data set was used for the
present study. Furthermore, solely the measured $^8$B-neutrino
spectrum~\cite{OGW00} has been included. The $\Delta m^2_{12}$-$\Delta
m^2_{13}$--plane 
is divided according to the 2-flavor 
case in three subregions where the two oscillation-branches
$\nu_1\!\leftrightarrow\!\nu_2$ and $\nu_1\!\leftrightarrow\!\nu_3$ are either
both of matter type, or both of vacuum type, or one of matter and the second
of vacuum nature. 

\subsubsection{Oscillations in the MSW mass regime} \label{purmsw}

\begin{figure}[ht]
\hbox to\hsize{\hss\epsfxsize=8.6cm\epsfbox{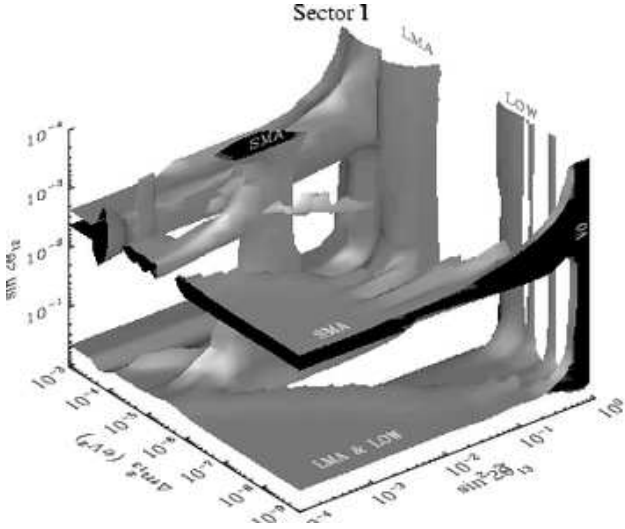}\hss}
\vspace*{4mm}
\caption{Shaded volume representing the projection of the region with
$\chi^2\!\le\!26.5$ into the 3-dimensional subspace of
$\sin^2\!2\Theta_{12}$-$\Delta m^2_{13}$-$\sin^2\!2\tilde{\Theta}_{13}$ for
MSW-transitions in the $\nu_1\!\leftrightarrow\!\nu_2$ and $\nu_1\!\leftrightarrow\!\nu_3$ system
(\mbox{\rm sector {\small I}}). The allowed values are located within the light shaded
surroundings. The position of the classical 2-flavor 
solutions are also shown, which may be in the $1\!\leftrightarrow\!2$
as well as in the $1\!\leftrightarrow\!3$ system.
\label{msw3}} 
\end{figure}

First \mbox{\rm sector {\small I}} is investigated, where $\nu_1\!\leftrightarrow\!\nu_2$
and $\nu_1\!\leftrightarrow\!\nu_3$ both may undergo resonant MSW-transition,
i.e.~$10^{-9}\le\Delta m^2_{12},\Delta m^2_{13}\le 10^{-3}\,{\rm eV^2}$. 
In Fig.~\ref{msw3} the projection of the region with $\chi^2\!\le\!26.5$ on the
3-dimensional subspace $\sin^2\!2\Theta_{12}$-$\Delta
m^2_{13}$-$\sin^22\tilde{\Theta}_{13}$ is shown. The allowed 
parameters are  located within the grey-shaded surroundings. 

\begin{figure}[b]
\hbox to\hsize{\hss\epsfxsize=8.6cm\epsfbox{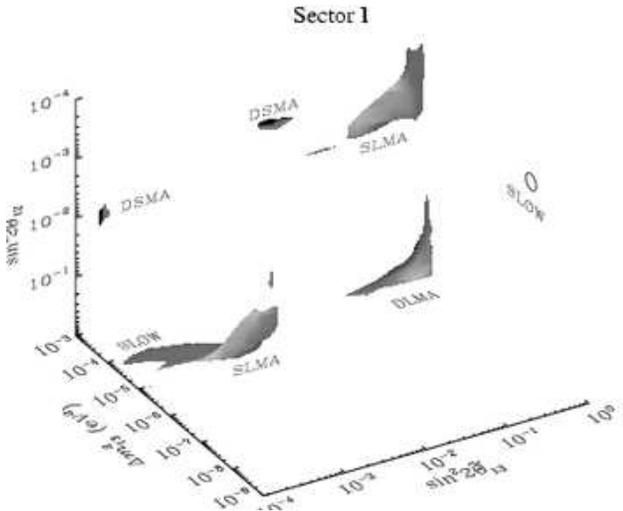}\hss}
\vspace*{4mm}
\caption{Zoom into the shaded volume of Fig.~\ref{msw3} showing the 
position of the SLMA-, DSMA-, DLMA-, 
and SLOW-solutions. The shaded bodies enclose the regions with
$\chi^2\!\le\!15.4$. The circle in the $\sin^22\Theta_{12}$-$\Delta
m^2_{13}$-plane at $\sin^22\tilde{\Theta}_{13}=1$ shows the position of the
tiny solution area of SLOW.\label{msw3_cl}}  
\end{figure}

The pure 2-flavor $\nu_1\!\leftrightarrow\!\nu_3$ oscillations where the mixing into
$\nu_{2}$ is negligible are given by vertical structures.
For instance, the LMA- and LOW-solutions are represented by the
half-pillars at $\sin^22\tilde{\Theta}_{13}=1$ and $\Delta
m^2_{13}\approx5\times10^{-5}$ and $10^{-7}\,{\rm eV^2}$. The
horizontal planes provide the solutions independent of $\nu_3$, and
thus are 2-flavor oscillations in the $1\!\leftrightarrow\!2$ system, where the
influence from $\Delta m^2_{12}$ is not 
visible. However,
the symmetry of \mbox{\rm sector {\small I}} in the exchange of the indices 2 and~3 implies
that they must be 
equivalent to solutions independent of $\nu_2$. Therefore, the 
plane around $\sin^22\Theta_{12}\approx1$ is
equivalent to the pillar-like structures at
$\sin^22\tilde{\Theta}_{13}=1$ and can thus be identified 
as an overlap of mainly LMA- and LOW-solutions. Similarly the plane at
$\sin^22\Theta_{12}\approx3\times10^{-3}$ represents the SMA-solution. The
3-flavor solutions with the 
smallest $\chi^2$-values are sited near the intersection regions of
the horizontal planes and vertical objects, which means that they are
at least slightly more probable than the pure 2-flavor solutions
and involve indeed all three flavors.

In Fig.~\ref{msw3_cl} the regions with $\chi^2\!\le\!15.4$ of the present
survey are shown. 
Due to the symmetry in $\nu_2\!\leftrightarrow\!\nu_3$ and the overlap of
solution islands in this projection the 6 basic regions in
this figure belong only to four distinct solutions. Two of them can
be identified as ``double'' SMA- resp.~LMA-solutions, i.e.~the same
kind of solution in $\nu_1\!\leftrightarrow\!\nu_2$
and $\nu_1\!\leftrightarrow\!\nu_3$ (DSMA resp.~DLMA). In addition, two solutions
are combinations of a SMA-solution in $\nu_1\!\leftrightarrow\!\nu_2$ 
($\sin^22\Theta_{12}\lesssim 10^{-3}$) and a LMA
resp.~LOW in $\nu_1\!\leftrightarrow\!\nu_3$ ($\sin^22\tilde{\Theta}_{13}\gtrsim
10^{-1}$ and $\Delta m^2_{13}\approx 10^{-4}$ resp.~$10^{-7}\,{\rm
eV^2}$.), therefore 
denoted SLMA and SLOW. The extension of the SLOW-solution in the
$\sin^22\Theta_{12}$-$\Delta m^2_{13}$-plane at
$\sin^22\tilde{\Theta}_{13}\approx1$ is very small
and thus its position is marked by a small circle.
The assignment of $\nu_1\!\leftrightarrow\!\nu_2$ to SMA and $\nu_1\!\leftrightarrow\!\nu_3$
to LMA resp.~LOW is ambiguous and could also be chosen vice
versa. This is reflected by the second appearance of the SLOW- and
SLMA-solutions in Fig.~\ref{msw3_cl} at $\sin^22\Theta_{12}\gtrsim10^{-1}$
and $\sin^22\tilde{\Theta}_{13}\lesssim10^{-3}$. 
The respective solution islands merge in this projection.

\begin{figure}[hb]
\hbox to\hsize{\hss\epsfxsize=8.6cm\epsfbox{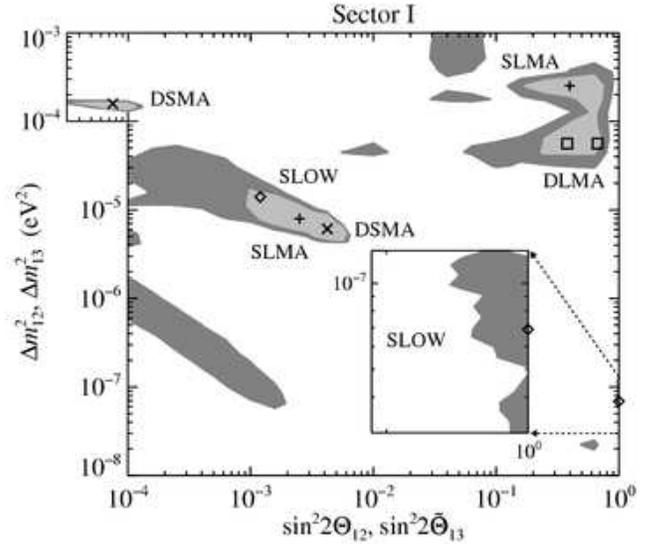}\hss}
\vspace*{2mm}
\caption{Projection of the DSMA-, DLMA-, SLMA-, and SLOW-solutions onto the
$\Delta m^2_{12}$-$\sin^22\Theta_{12}$-- or the
$\Delta m^2_{13}$-$\sin^22\tilde{\Theta}_{13}$--plane. The dark and bright
shaded areas show the regions of $\chi^2\!\le\!15.6$ and
13.9. Following the assignment of
Table \ref{bestmsw3}, the $\chi^2\!\le\!13.9$--region of  
the DSMA-, SLMA-, and SLOW-solutions merge in the projection onto the
$\Delta m^2_{12}$-$\sin^22\Theta_{12}$--plane, but 
are well separated in the $\Delta m^2_{13}$-$\sin^22\tilde{\Theta}_{13}$--plane
(see also Fig.~\ref{msw3_cl}). The region with
$\sin^22\tilde{\Theta}_{13}<10^{-4}$ has been scanned separately only
near the DSMA-solution ($10^{-6}<\Delta m^2_{12}<10^{-4}\,{\rm eV^2}$
and $10^{-4}<\sin^22\Theta_{12}<10^{-2}$).\label{slma}}
\end{figure}

In Fig.~\ref{slma} the projection of the DSMA-, DLMA-, SLMA-, and
SLOW-solutions into the typical 2-flavor planes are shown. One of
the areas labelled, for instance, SLMA has to be identified with
oscillations in $\nu_1\!\leftrightarrow\!\nu_2$ and the other with
$\nu_1$-$\nu_3$--mixing. The
oscillation parameters are provided in Table~\ref{bestmsw3} together with
the $\chi^2$-values. Since two more
parameters are adjusted in the 3-neutrino as compared to the
2-neutrino case, the number of degrees-of-freedom reduces from 26 to 24.

\begin{table*}[ht]
\protect\caption{The most favored solutions within the 3-flavor
oscillation scenario 
and the respective $\chi^2$-values. The second column indicates the
sector in the flavor space, where the solution has been found (see
also Fig.~\ref{spher}). $\Delta m^2_{ij}$ is in
units of eV$^2$ and $s^2_{ij}$ abbreviates here
$\sin^22\Theta_{ij}$. In \mbox{\rm sector {\small I}} $\sin^22\tilde{\Theta}_{13}$ is defined
by \mbox{Eq.~(\ref{angsym})}, analogously in 
\mbox{\rm sector {\small II}} by \mbox{Eq.~(\ref{angII})}. $\chi^2_R$, $\chi^2_e$, $\chi^2_Z$ and
$\chi^2_A$ are the individual contributions to $\chi^2_{\rm tot}$ as
defined by Eqs.~(\ref{chir}) and (\ref{chie}).
The number of d.o.f.~is 24.\label{bestmsw3}}
\center\begin{tabular}{cclclcddddd}
&  & \multicolumn{1}{c}{$\Delta m^2_{12}$} &
\multicolumn{1}{c}{$s^2_{12}$} & \multicolumn{1}{c}{$\Delta m^2_{13}$} & 
\multicolumn{1}{c}{$\tilde{s}^2_{13}$} & \multicolumn{1}{c}{$\chi^2_R$}
& \multicolumn{1}{c}{$\chi^2_e$} & \multicolumn{1}{c}{$\chi^2_Z$} &
\multicolumn{1}{c}{$\chi^2_A$} & 
\multicolumn{1}{c}{$\chi^2_{\rm tot}$/d.o.f.} \\ \hline   
DSMA & {I} & $6.1\times10^{-6}$ & $4.2\times10^{-3}$ & $1.6\times10^{-5}$ & $7.5\times10^{-5}$ & 0.6 & 8.2 & 1.9 & 1.2 & 0.50 \\
DLMA & {I} & $5.6\times10^{-5}$ & 0.67 & $5.6\times10^{-5}$ & 0.38 & 0.9 & 8.5 & 2.0 & 1.3 & 0.53 \\
SLMA & {I} & $7.9\times10^{-6}$ & $2.5\times10^{-3}$ & $2.5\times10^{-4}$ & 0.40 & 0.6 & 8.5 & 1.8 & 1.1 & 0.50 \\
SLOW & {I} & $1.4\times10^{-5}$ & $1.2\times10^{-3}$ & $6.9\times10^{-8}\!$& 1.00 &  1.8 & 8.7 & 2.1 & 2.1 & 0.61\\
DVO & {I} & $5.5\times10^{-11}\!$& 0.64 &$2.9\times10^{-10}\!$& 0.60 & 3.5 & 5.7 & 1.9 & 1.5 & 0.53\\
DVO'& {II} & $1.2\times10^{-10}\!$& 0.58 &$7.3\times10^{-11}\!$& 0.79 & 2.3 &6.8 & 1.9 & 1.5 & 0.52 \\
SVO & {I} & $2.7\times10^{-10}\!$& 0.51  & $1.0\times10^{-5}$ & $2.5\times10^{-3}$ & 1.6 &
5.9 & 1.8 & 0.9 & 0.43 \\
LVO & {I} & $4.1\times10^{-11}\!$& 0.70 & $1.7\times10^{-4}$ & 0.40 &
1.4 & 8.1 & 2.0 & 2.0 & 0.56 \\\hline
d.o.f.&&&&&&3 & 17 & 5 & 3 &\\
\end{tabular}
\end{table*}

\begin{figure}[b]
\hbox to\hsize{\hss\epsfxsize=8.6cm\epsfbox{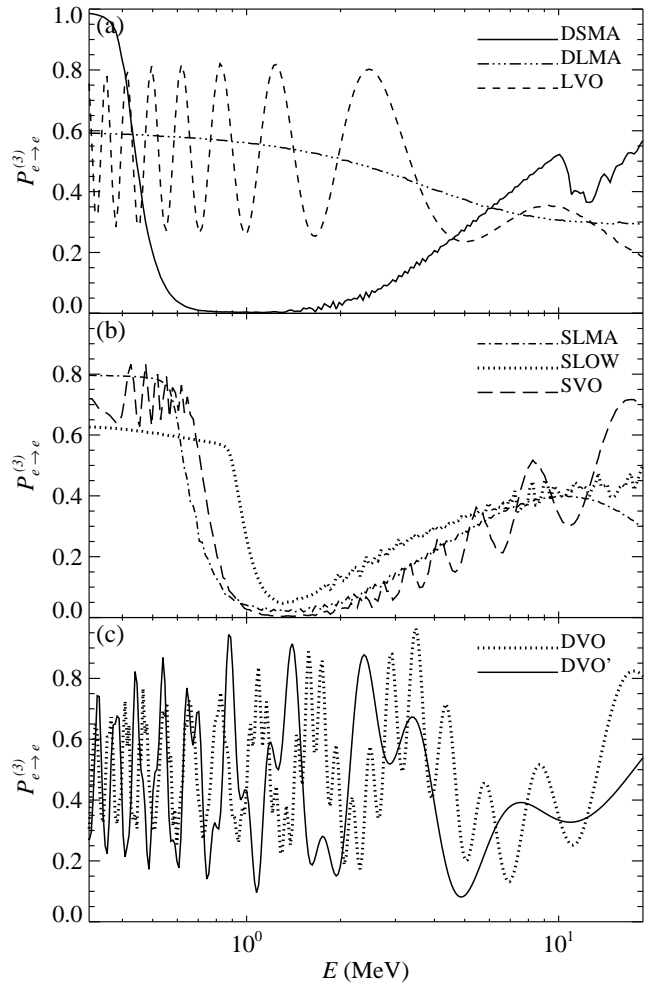}\hss}
\caption{The annually averaged survival probability of $^8$B
electron-neutrinos detected on Earth for the deduced 3-flavor
solutions.\label{p_e3}} 
\end{figure}

The mixing angle of the second small mixing-angle branch in the
DSMA-solution at $\sin^22\tilde{\Theta}_{13}$ is about two order of
magnitude smaller than the usual 2-flavor 
SMA-solution (compare Figs.~\ref{slma} and \ref{msw2}). This second SMA-branch
causes an additional resonance at $E\approx10\,{\rm MeV}$ (see
Fig.~\ref{p_e3}), which enables a better fit to the recoil-electron
energy spectrum and the event rates (compare Tables \ref{totmsw2} and
\ref{bestmsw3}). Thus a solution almost as good as the LMA-solution
could be obtained ($\chi^2_{\rm tot}/{\rm d.o.f.} = 0.5$).

The DLMA- and SLMA-solutions have almost the same values for $\chi^2_R$,
$\chi^2_e$, $\chi^2_Z$ and $\chi^2_A$ as the LMA-solution
(cp.~Tables \ref{totmsw2} and 
\ref{bestmsw3}). Since thus no improvement in explaining the solar
neutrino measurements could be achieved by combining the 
LMA-solution with an additional oscillation in the
$\nu_1\!\leftrightarrow\!\nu_3$-system, the 2-flavor LMA-solution itself
is a favored solution in the 3-flavor analysis, too

In the SLOW-solution the advantages of the 2-flavor SMA- and
LOW-solutions are united. While with the SMA-solution the rates can be
reproduced very well (see Table~\ref{bestmsw2}), the LOW-solution
provides good fits to the energy spectrum, the annual and the zenith
angle variations (Table~\ref{totmsw2}). 

The best-fit value of the SLOW-solution for
$\sin^22\tilde{\Theta}_{13}$ is equal to 1, and thus at the border of
\mbox{\rm sector {\small I}} to \mbox{\rm sector {\small III}}. Exchanging the assignment of the indices 2 and 3
transfers the SLOW-solution to the border between \mbox{\rm sector {\small I}} and
{II}. In the latter sector,  where
$\pi/4\le\sin\Theta_{12}\le\pi/2$ and
$\tan\Theta_{13}\le\sin\Theta_{12}$ a survey has been performed,
too. However no new solutions could be 
identified; solely, the ``foothills'' of the SLOW-solution into this
section have been found. Hence, in the mass regime with
$10^{-9}\le\Delta m^2_{12},\Delta m^2_{13}\le 10^{-3}\,{\rm eV^2}$ the 
region with $\pi/4\le\sin\Theta_{12}\le\pi/2$ does not provide new
solutions, which is similar to the findings in the 2-flavor case,
where also no additional solutions have been detected in this
parameter range. 
\begin{figure}[b]
\hbox to\hsize{\hss\epsfxsize=8.6cm\epsfbox{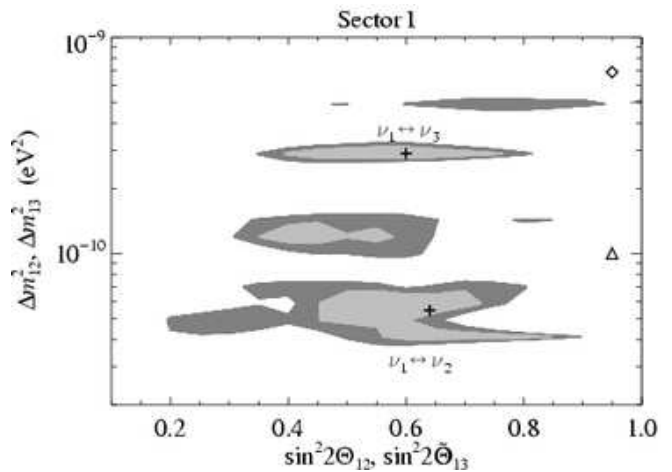}\hss}
\vspace*{2mm}
\caption{Projection of the DVO-solution on the
$\Delta m^2_{12}$-$\sin^22\Theta_{12}$-- resp.~$\Delta
m^2_{13}$-$\sin^22\tilde{\Theta}_{13}$--plane. The $\chi^2$-levels of 
the regions agree with those of Fig.~\ref{slma} and the ``$+$''-symbols
denote the best-fit values.  
The best-fit value of the 2-flavor VO-solution reproducing solely
the event rates is marked by $\triangle$, while the position of the
2-flavor  VO-solution in the complete analysis is shown
by $\diamond$. \label{vac3}}
\end{figure}

\subsubsection{Three-flavor vacuum oscillations}

\begin{figure}
\hbox to\hsize{\hss\epsfxsize=8.6cm\epsfbox{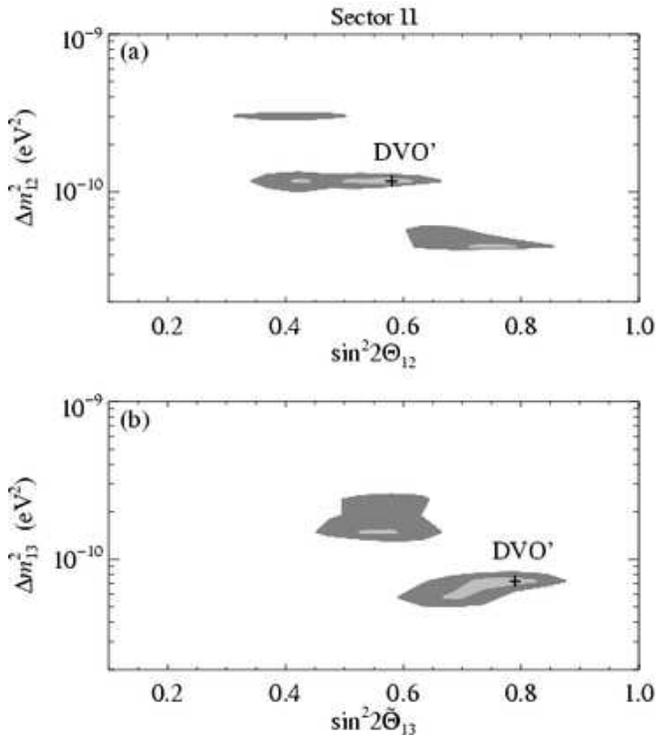}\hss}
\vspace*{2mm}
\caption{Projection of the DVO'-solution on the
$\Delta m^2_{12}$-$\sin^22\Theta_{12}$-- and $\Delta
m^2_{13}$-$\sin^22\tilde{\Theta}_{13}$--plane. The $\chi^2$-levels of 
the regions agree with those of Fig.~\ref{vac3}. 
\label{vac3_2}}
\end{figure}

In the case when both mass-squared differences are in the
vacuum-oscillation regime
($\Delta m^2_{12}$,$\Delta m^2_{13}\le 10^{-9}\,{\rm eV^2}$) two
minima could be found, denoted DVO and DVO'. While DVO is
located in \mbox{\rm sector {\small I}} (Fig.~\ref{vac3}),
i.e.~$0\le\Theta_{12}\le\pi/4$, DVO' has 
been found in \mbox{\rm sector {\small II}}
(Fig.~\ref{vac3_2}), where $\pi/4<\Theta_{12}\le\pi/2$.
Thus although they seem to have very similar mixing parameters (Table
\ref{bestmsw3}), there 
are really distinct solutions with different properties (see Fig.~\ref{p_e3}). Note that, while in
\mbox{\rm sector {\small I}} the assignment of the indices can be exchanged, this is not
possible in \mbox{\rm sector {\small II}}. If an exchange of $\nu_2$ and $\nu_3$ in DVO' is
desired, the mixing angles have to be transformed appropriately to lie
finally in \mbox{\rm sector {\small III}}.

With the DVO- and DVO'-solutions a slightly better fit to the
recoil-electron  energy spectrum can be achieved compared to the
DSMA-, DLMA-, and SLMA-solutions (Table~\ref{bestmsw3}). However,
because of the event rates being reproduced worse, the
$\chi^2_{\rm tot}$-values of DVO and DVO' are almost equal to the values of 
DSMA, DLMA, and SLMA.

Compared to the 2-flavor VO-solution an improved fit to the event
rates has been obtained with the 3-flavor vacuum-oscillation solutions, 
but still the event rates are fitted barely
acceptably. The DVO- and
DVO'-solutions show a compromise between the VO-solution ($\triangle$
in Fig.~\ref{vac3}) 
obtained by fitting solely the event-rate data and the VO-solution of
the complete analysis ($\diamond$ in same figure). The mixing angles
for the 3-flavor vacuum solution are only about one half of the usual
2-flavor VO-solution, as $\nu_e$ is now oscillating nearly equally
strong into \emph{two} other flavors. 


\begin{figure}[t]
\hbox to\hsize{\hss\epsfxsize=8.6cm\epsfbox{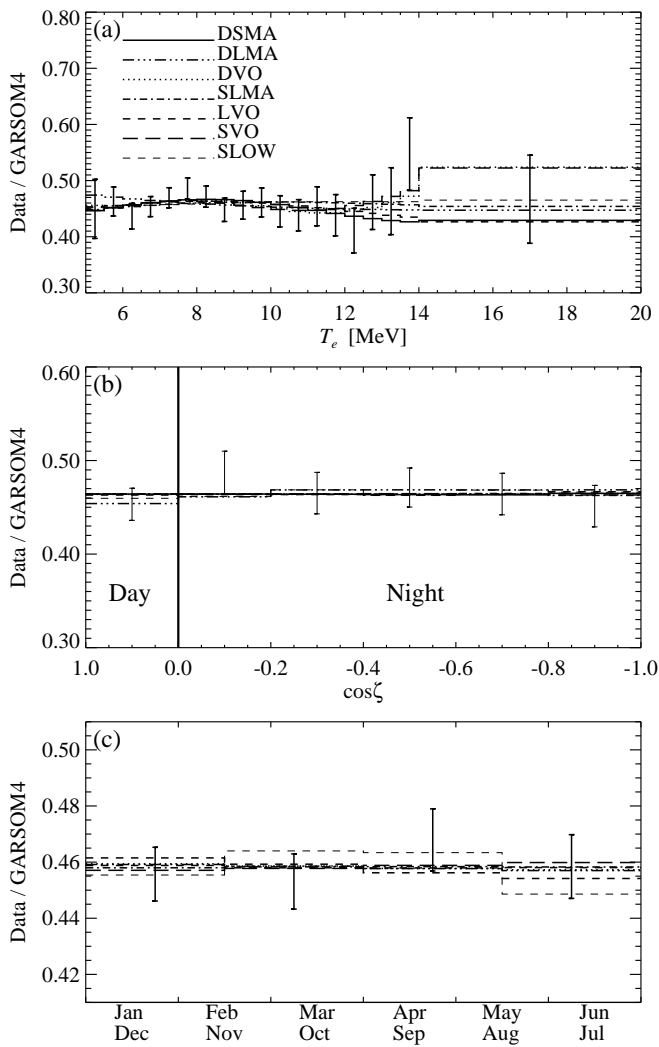}\hss}
\caption{Recoil-electron energy spectrum (a), zenith-angle ($\zeta$)
variation (b), 
and seasonal dependence (c) of the best fit 3-flavor solutions
including all available solar neutrino data compared with the
measurements of Super-Kamiokande. The experimental data agree with
Fig.~\protect\ref{esp_msw2} using the measured $^8$B-neutrino spectrum
\protect\cite{OGW00}.\label{esp_msw3}}
\end{figure}

\subsubsection{Mixed vacuum and MSW oscillations}

The combination of oscillation between two neutrino flavors in the
MSW mass-regime and an additional vacuum oscillation into the third flavor
is investigated completing the mass ranges which have not been
covered by the previous sections. This case has been examined in
\cite{Liu97} for $10^{-7}\le\Delta m^2_{13}\le10^{-4}{\rm eV^2}$ and
$10^{-12}\le\Delta m^2_{12}\le10^{-9}\,{\rm eV^2}$. However, in that
analysis only the event rates have been included, but not the recoil-electron
energy spectrum, the zenith-angle dependence nor the annual variation data
recorded by Super-Kamiokande. Furthermore, the Earth-regeneration
effect has been neglected. Therefore, this case is
reinvestigated including all the available data and calculating the
electron-neutrino survival probabilities fully consistently including
the Earth effect like in the previous sections. 

In this survey $\Delta m^2_{13}$ is taken to be in the mass range of the
MSW-solutions ($10^{-9}\le\Delta m^2_{13}\le10^{-3}\,{\rm eV^2}$)
and $\Delta m^2_{12}$ in the vacuum oscillation area
($10^{-12}\le\Delta m^2_{12}\le10^{-9}\,{\rm eV^2}$). With these
conventions about $\Delta m^2_{13}$, two possibilities are 
conceivable for the $\nu_1\!\leftrightarrow\!\nu_3$ system, a resonant and a
non-resonant oscillation. Since the masses are well separated, the
condition for a resonance obtained in the 2-flavor case can be applied
(Eq.~\ref{reson})
\[
N_{\rm res} = \frac{\Delta m_{13}^2/2E}{\sqrt{2} G_{\rm
F}}\cos\!2\Theta_{13} > 0.
\]
Thus, an MSW flavor transition is only possible,
if the mixing angle $\Theta_{13}$ is less than $\pi/4$. 
In the $\nu_1\!\leftrightarrow\!\nu_2$ system only
pure vacuum oscillations occur, and thus $\Theta_{12}\le\pi/4$.
Since the solution may be a combination of a vacuum oscillation in the
$1\!\leftrightarrow\!2$ and a resonant resp.~non-resonant oscillation in the
$1\!\leftrightarrow\!3$ branch, the complete essential parameter 
space for $10^{-9}\le\Delta m^2_{13}\le10^{-3}\,{\rm eV^2}$ and 
$10^{-12}\le\Delta m^2_{12}\le10^{-9}\,{\rm eV^2}$ is therefore
covered by $\Theta_{12}\le\pi/4$ and $\Theta_{13}\le\pi/2$.

\begin{figure}[t]
\hbox to\hsize{\hss\epsfxsize=8.6cm\epsfbox{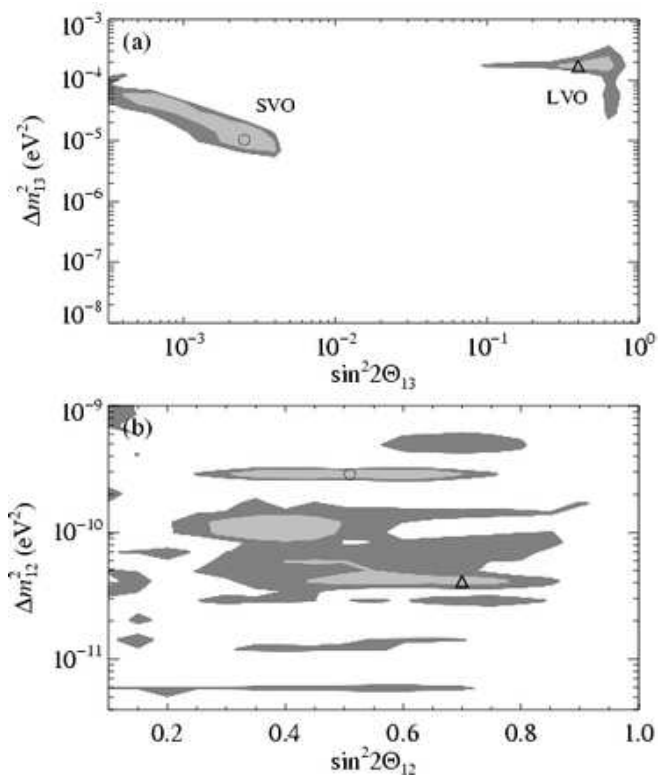}\hss}
\vspace*{2mm}
\caption{Projections of the SVO- and LVO-solutions onto the $\Delta
m^2_{12}$-$\sin^22\Theta_{12}$-- and $\Delta
m^2_{13}$-$\sin^22\Theta_{13}$--plane.
The $\chi^2$-levels are chosen to be 13.9 (bright) and
15.6 (dark shaded) like in the previous figures. The circles mark
the best-fit values of the SVO-solution, the triangles the one of the
LVO-solution. ($\Theta_{12},\Theta_{13}\le\pi/4$)
\label{svac}}
\end{figure}

Basically two minima could be found in the parameter space of this
subsurvey, which are  
combinations of a SMA- resp.~a LMA-solution in $\nu_1\!\leftrightarrow\!\nu_3$
($\Theta_{13}\le\pi/4$) and an additional
VO-mixing in $\nu_1\!\leftrightarrow\!\nu_2$, therefore denoted SVO and LVO.
The SVO-solution is similar to the solution ``B'' found in
\cite{Liu97}. Interestingly, the vicinity of the SVO-solution has been
explored intensively in~\cite{Bab98}, whereas the
motivation was different: The mixing parameters of this solution agree with
the neutrino properties predicted by a grand unification (GU) theory,
where the neutrino masses are caused by the seesaw mechanism with the
mass of the heaviest right-handed neutrino being of the order of the
GU-scale. 

The fit of the LVO-solution to the data is worse than the pure
2-flavor LMA-solution supporting the result of part 1 of this
section, that the LMA-solution cannot be improved by including
a third oscillation. However, the mass-squared difference of the
LMA-branch in the LVO- as well as in the SLMA-solution ($\Delta
m^2_{13}\approx2\times10^{-4}\,{\rm eV^2}$) is
almost one order of magnitude higher than that of the LMA-solution
itself. The implication of this fact for the combined analysis of
solar and atmospheric neutrino problem will be discussed further in
the final section. 

The SVO-solution is presently the favored solution to 
the solar neutrino problem, in contrast to all other 3-flavor
solutions even better than the 2-flavor LMA-solution. SVO combines
the merit of the 
VO-solution to explain the energy spectrum and the property of
the SMA-solution to reproduce the event rates (Table~\ref{bestmsw2}).
Interestingly, this solution has also been found to be the best explanation of
the solar neutrino problem, using the older 504- or 708-days data sets of
Super-Kamiokande.

\section{Prospects for future experiments}\label{future}
Future neutrino experiments like SNO or Borexino will provide further
information to determine the correct solution to the solar neutrino problem.
In the following the ability of these detectors to discriminate the
various 2- and 3-flavor solutions will be discussed.

\subsection{Borexino}
In Fig.~\ref{bor_msw3} the expected recoil-electron energy
spectrum in Borexino of the 2- and 3-flavor solutions is shown.
The shades and hatched areas indicate the uncertainties in the mixing
parameters of the four 2-flavor solutions. These areas were obtained
by allowing 
the mixing parameters to vary within the 10\% confidence
region of the $\Delta m^2$-$\sin^2\!2\Theta$--plane (see
Fig.~\ref{mswfull2}). 

\begin{figure}[hb]
\hbox to\hsize{\hss\epsfxsize=8.6cm\epsfbox{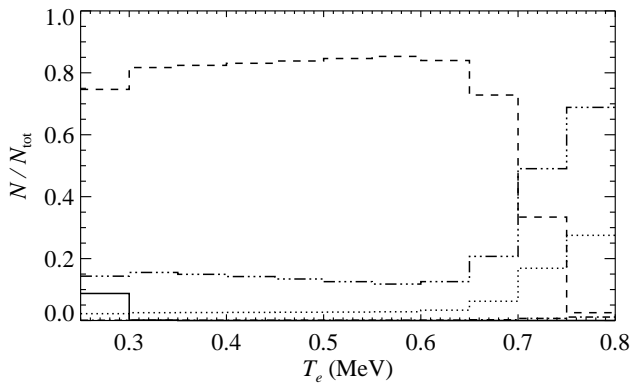}\hss}
\caption{The contribution of different neutrino types to the total
event rate in Borexino in the case of no oscillations. The neutrino
flux was derived from GARSOM4. The linestyles correspond to pp-
(solid), pep- (dotted), $^7$Be- (dashed), CNO- (dash-dot-dot-dotted),
and $^8$B-neutrinos (dash-dotted).\label{bor_cont}}
\end{figure}

\begin{figure}[b]
\hbox to\hsize{\hss\epsfxsize=8.6cm\epsfbox{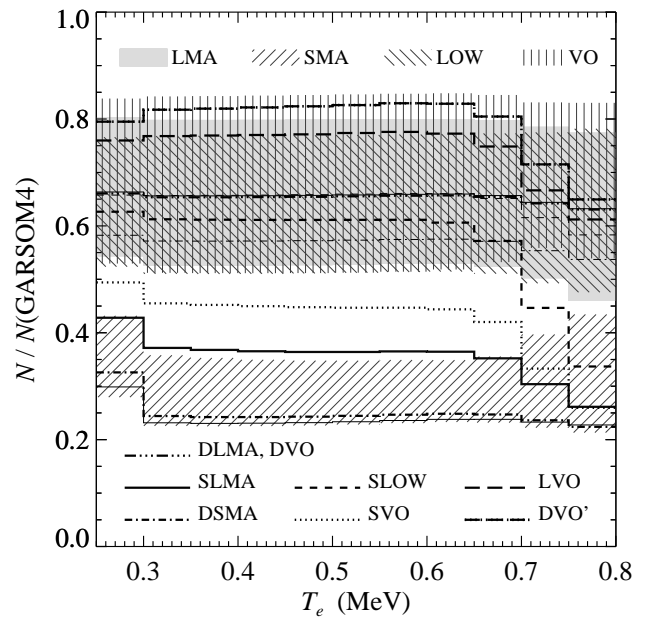}\hss}
\caption{Expected recoil-electron energy spectrum in Borexino for the
best-fit 2- and 3-flavor solutions. The shaded and hatched areas have been
obtained by allowing the mixing parameters to vary within the
10\%-region of the 2-flavor LMA resp.~SMA-solution shown in
Fig.~\ref{mswfull2}. The thin lines represent the best-fit 2-flavor
solutions in the respective area (solid: SMA resp.~LMA,
dash-dotted: LOW, dashed line: VO).\label{bor_msw3}}
\end{figure}

While in the presently operating detectors, the $^7$Be-neutrinos
contribute, if at all, only a small portion to the total event rate,
Borexino is going to measure basically these	
neutrinos (Fig.~\ref{bor_cont}). In addition, the highest
energy bin of the recoil-electron spectrum ($T_e\ge0.75\,{\rm
MeV}$) is sensitive to the neutrinos
of the CNO-cycle and of the pep-reaction. Borexino may also measure
the high energy part of the pp-neutrino spectrum, as these may
contribute about 9\% to the total 
rate in the lowest energy bin ($T_e\le0.3\,{\rm MeV}$). 

In the case of the SMA-solution Borexino should detect
only about 25\% of the event rate as predicted by GARSOM4, while in the
LMA-case about 65\% are expected (Fig.~\ref{bor_msw3}). Actually, for
the SMA-solution the rate is basically due to the interaction of the
$\nu_\mu$ (or $\nu_\tau$) with the detector, as in this solution the
$^7$Be-neutrinos almost totally consist of $\nu_\mu$-flavor. This
leads to an unmistakable signal of the SMA compared to 
the LMA-solution: An excess in the lowest energy bin of about 30\%
compared to the other bins caused  by the 
pp-neutrinos. With the resonance in the SMA-solution being at about
0.3\,MeV, pp-neutrinos do, unlike $^7$Be- or CNO-neutrinos,
not undergo an MSW-transition. Thus, with good  
statistics in the low 
energy bins SMA could be clearly distinguished from LMA. 

One of the main advantages of this detector is the ability to monitor 
seasonal variations in the neutrino signal --- the vacuum survival
probability is modified by the eccentric orbit of
the Earth (Eq.~\ref{Pvac2}).
Since the $^7$Be-neutrinos are in contrast to $^8$B-neutrinos emitted to
90\% in a monoenergetic line at 0.862\,MeV the signal in the
detector is not smeared out like in Super-Kamiokande. A detailed analysis about
the sensitivity of Borexino to detect vacuum oscillations is given in
\cite{Gou99}. 

\begin{figure}[ht]
\hbox to\hsize{\hss\epsfxsize=8.6cm\epsfbox{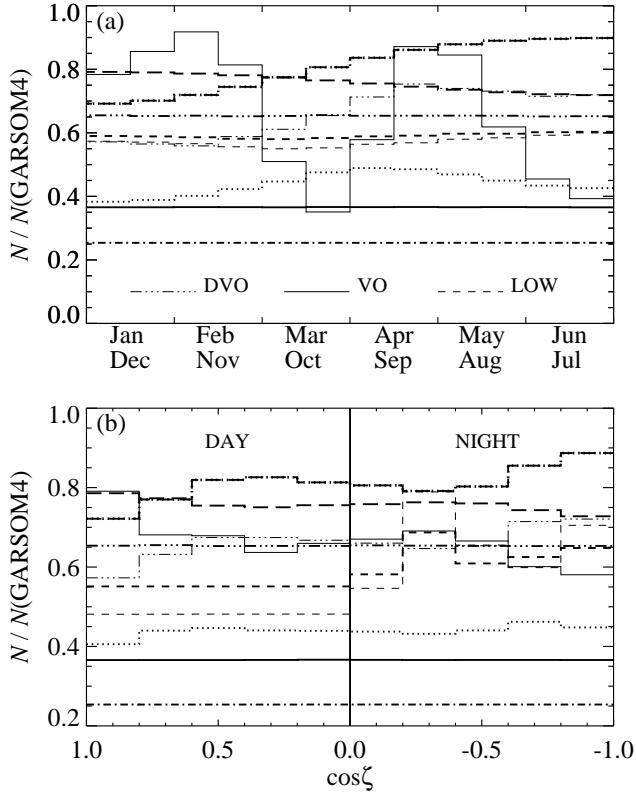}\hss}
\vspace*{3mm}
\caption{Annual and zenith-angle variations of the recoil electrons in
Borexino 
expected for the best-fit 
3-flavor solutions. The thick lines correspond to same solutions as
in Fig.~\ref{bor_msw3}. In the seasonal variation (b) the flux
changes due to the eccentricity of the Earth's orbit have been
subtracted.\label{bor2_msw3}}   
\end{figure}

The sensitivity of the $^7$Be-neutrino
flux in the VO-regime is also visible in the zenith-angle dependence
of the event rate in Borexino (thin solid line in Fig.~\ref{bor2_msw3}(b)). 
In particular, the rates of those zenith angles, under which the Sun
appears only around the winter or summer solstice ($|\cos\zeta|\gtrsim0.6$),
are influenced by the modulation of the $^7$Be-neutrino signal with
the Sun-Earth distance. Note that unlike the
Earth regeneration effect, which in the MSW mass-range can lead to a
variation of the event rate with the zenith angle only during the
night (see thin dash-dotted line in Fig.~\ref{bor2_msw3}(b)
representing the LOW-solution),
for VO-solutions 
a variation in the signal over the whole day is apparent. The
magnitude of  the zenith-angle variation however is smaller, as each
zenith-angle bin is an average over many positions of the Earth in its
solar orbit. Therefore, the seasonal variation of the event rates are
more sensitive to detect VO-solutions as the zenith-angle dependence.
For the best-fit mixing parameters of the MSW-solutions SMA and LMA 
no Earth regeneration takes place for the energy of the
$^7$Be-neutrinos, and thus no zenith-angle variation of the event
rates in Borexino is expected for these cases. In case of the
LOW-solution, however, day-night variations should be measured (thin
dash-dotted line in Fig.~\ref{bor2_msw3}(b)).

In summary for the 2-flavor case, a total rate of about $30\pm5\%$
compared to GARSOM4 is predicted 
for the SMA-solution, and about $65\pm15\%$ for the \mbox{LMA-,} LOW-, and
VO-solutions. The latter three can be disentangles by using the
zenith-angle and seasonal data. Great seasonal variations are
expected for the VO-solution, strong zenith-angle dependence during
the night for LOW, and a constant neutrino flux during the whole year for the
LMA-solution, provided that the influence of the eccentric Earth orbit
has been removed. Note, that also for the SMA-solution neither zenith-angle
nor seasonal variations are expected.

The expected energy, zenith-angle, and seasonal dependences of the
recoil electrons in 
Borexino for the best-fit values of the eight 3-flavor solutions
are shown in  
Figs.~\ref{bor_msw3} and \ref{bor2_msw3}, too. The DSMA and DLMA
cause very similar signals as their 2-flavor counterparts SMA
resp.~LMA, and are therefore difficult to
disentangle from the latter in any kind of the Borexino data. 

But the SLOW-solution leads to a clear signature in 
Borexino compared to the 2-flavor solutions. It 
has a strongly reduced event rate of almost 50\% in the high-energy
bins of Borexino, which are not observed in any other 
solution. The reason for this is the relatively high resonance energy
of about 1\,MeV (cf.~Fig.~\ref{p_e3}), which causes a strong
suppression of electron-type CNO- and pep-neutrinos, but a weaker
reduction of $^7$Be electron-neutrinos. By way of contrast, the
resonances of the SLMA- and SVO-solutions are about 0.8\,MeV
(Fig.~\ref{p_e3}), and thus, the $^7$Be electron-neutrino are
diminished stronger than in the SLOW-solution, which according to
Fig.~\ref{bor_cont} yields smaller rates below 0.7\,MeV.

In addition, the SLOW-solution leads to a zenith-angle
dependent signal during the night due to the Earth regeneration effect
(Fig.~\ref{bor2_msw3}(b)) caused by the LOW-branch of this
solution. Thus, a large day-night variation measured by Borexino
would strongly favor LOW or SLOW. By additional data from the
energy spectrum these solution should then be distinguishable from
each other.

The mean event rates expected for the SLMA- and SVO-solutions are
within a region, where all other solutions are disfavored. Since for
the SVO-solution also annual variations are expected from its
VO-branch (see Fig.~\ref{bor2_msw3}(a)), both solutions may be
disentangled from each other and from all other solutions.

The properties of the pure 3-flavor vacuum-oscillation solutions DVO
and DVO' are very sensitive to the exact mixing
parameters.
Even a very small variation in the mixing parameters of these
solutions may produce a recoil-electron
energy spectrum or seasonal changes which resembles that of the
other or the LVO-solution.
Hence a discrimination between DVO, DVO', and LVO will almost be
impossible with Borexino.

These solutions are also difficult to distinguish from the
2-flavor VO-solution. However, the most probable value for $\Delta
m^2$ of the VO-solution is almost one order of magnitude larger than
the VO-branch(es) of DVO, \mbox{DVO'} or LVO (compare Tables~\ref{totmsw2} and
\ref{bestmsw3}) yielding a more frequently varying annual
signal for the VO-solution. Hence, measuring a weak
annually varying signal with only one maximum per half of the year
would at least provide
a strong hint, that the solution to the solar neutrino problem is
rather a 3-flavor DVO-, DVO'- or LVO-solution than a pure
2-flavor VO solution.

\subsection{SNO}
Recently, SNO has started to measure solar neutrinos in the same
energy window as Super-Kamiokande, however 
with a better energy resolution and the ability to discriminate the CC-
and NC-events. Since the total number of neutrinos does not change
by neutrino oscillations (only their flavor), the number of
NC-events has to agree with the no-oscillation scenario, i.e.~the
total number of solar neutrinos independent of their flavor can be
determined by this quantity\footnote{~Provided, no mixing with sterile
neutrinos occurs.}. Therefore, no   
departures in the recoil-deuterium energy spectrum, zenith-angle or
seasonal dependence from the standard (no-oscillation) values are
expected for NC-events. The merit of SNO is the measurement of the
ratio of CC- to NC-events, as with this quantity the
$\nu_e$-contribution to the total solar neutrino flux is connected. A
measured CC/NC-ratio smaller than the standard one unambiguously
proves that the solar neutrinos, initially $\nu_e$'s, are oscillating 
into another flavor during their flight from the solar interior to
the Earth. 
 
\begin{figure}[ht]
\hbox to\hsize{\hss\epsfxsize=8.6cm\epsfbox{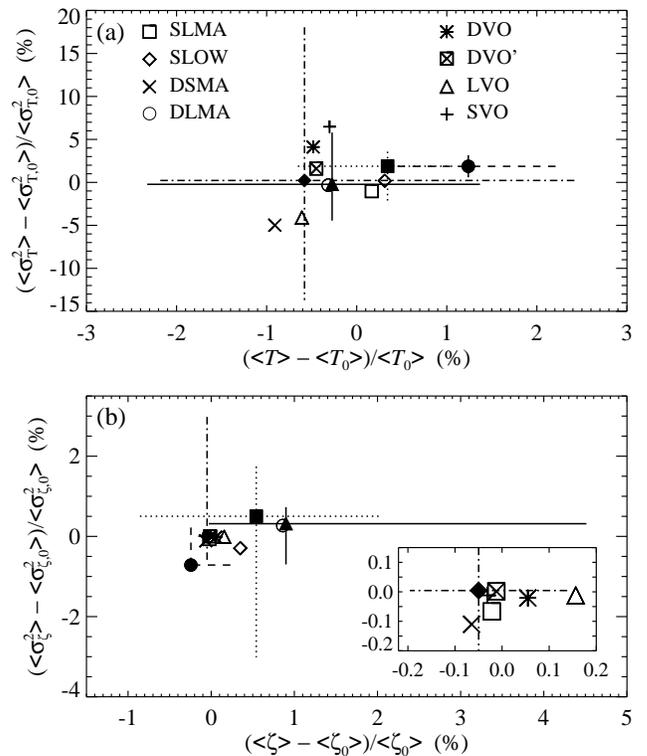}\hss}
\vspace*{2mm}
\caption{Predicted values for the moments of the recoil-deuterium
energy 
(upper panel) and  zenith-angle spectrum (lower panel) for
CC-events in SNO in case of the 3-flavor solutions. The filled
symbols represent the 2-flavor solutions LMA ($\blacktriangle$), SMA
($\bullet$), LOW ($\blacksquare$) and VO ($\blacklozenge$) of
Fig.~\ref{msw2}. The error bars were obtained in the same manner as
the hashed resp.~shaded areas in Fig.~\ref{bor_msw3}.\label{sno_msw3}}   
\end{figure}

If the excess in the high-energy bins of Super-Kamiokande, which was
apparent in the previous data published~\cite{SK99e}, reappears in the   
continuing measurements, the CC/NC-ratio in SNO will enable to
discriminate between an explanation by neutrino oscillations and by an
enhanced $hep$-neutrino 
flux, which also was under discussion~\cite{Bah99}. While for
the latter solution the $hep$-neutrinos would equally contribute to
the CC- and NC-events leading to a constant CC/NC-ratio, for
the oscillation scenario a similar excess like in Super-Kamiokande would be
also apparent in the CC/NC-ratio of SNO.  

To illustrate the expectations for the recoil-deuterium energy spectrum and
zenith-angle dependence of the CC-events in SNO, the first and second 
moments of these distributions are calculated (for definition see
\cite{BahK97}). The first
moment provides the mean value and the second a measure of the width
of the distribution. 

With very small annual or daily changes in the Super-Kamiokande detector,
such variations are neither expected for CC-events ($e$-$\nu_e$--scattering)
in SNO. Thus, for the zenith-angle moments in SNO only small
deviations from the standard value 
are predicted for the 2-flavor SMA-, LMA-, LOW-, and VO-solutions
(Fig.~\ref{sno_msw3}(b)). The largest deviations of about 4\% may be
measured in case of the LMA-solution being correct.

The error bars for these solutions in
Fig.~\ref{sno_msw3} were obtained in the same manner as for the
recoil-electron energy spectrum in Borexino. Since the best-fit values
are not very well separated and also the error bars strongly overlap
in the plane of the first two zenith-angle  
moments, it seems to be very difficult to disentangle the 2-flavor 
solution by forthcoming zenith-angle data of SNO. 
A similar behavior is expected for the moments of the recoil-deuterium 
energy spectrum~\cite{Bah97b} shown in
Fig.~\ref{sno_msw3}(a). However, since the  
energy resolution in SNO is better than in Super-Kamiokande
(Table~\ref{detpar}) more significant data can be expected to
disentangle the true solution to the solar neutrino problem.
A more detailed discussion about the properties of SNO to
discriminate the 2-flavor solutions can be found in~\cite{Bah00b}.

In the energy moments a large area is still allowed for the
VO-solution. In case Borexino does not record any seasonal variations
all solutions containing a VO-branch (e.g.~LVO), and in particular the
2-flavor VO-solution would hardly be correct. Deviation of
$< -5\%$ in the second energy moment would then favor the
DSMA-solution.

Nevertheless, from Fig.~\ref{sno_msw3} it becomes obvious that, only if
SNO is measuring a big deviation of the energy or zenith-angle moment
from the standard (no-oscillation) value, one of these solution can
be favored, as e.g.~$\gtrsim3\%$ in the first zenith angle further
supports LMA as correct solution to the solar neutrino problem.
But, since SNO may well detect no such deviations, we need additional
results from Borexino to exclude some of the presently possible
solutions.  

\section{Discussion}
The most favored solutions to the solar neutrino problem
in the scope of 2- and 3-flavor neutrino oscillations have been
investigated in this work.  
For this purpose the evolution of the neutrino state was followed
numerically from the solar interior, where the neutrinos are produced in
the fusion reactions, through Sun, space and
Earth to the neutrino detectors without falling back to 
analytical approximations (cf.~Appendix).

Eight new solutions in the 3-flavor case could be found, among
which SVO, SLMA, and DSMA are at least  
as probable as the presently favored 2-flavor LMA-solution. The DVO-,
\mbox{DVO'-,} LVO, DLMA-, and, SLOW-solutions, with which the measured event
rates, recoil-electron energy spectrum, zenith-angle dependence and annual
variation are reproduced slightly worse 
than with the other three 3-flavor solutions, are nevertheless still good
candidates as solution 
to the solar neutrino problem. The new
experiments Borexino and SNO, which recently have started to operate,
and improved statistics in Super-Kamiokande possibly  
will possibly enable to discriminate the various 2- and 3-flavor
solutions. Basically, with Borexino the vacuum-oscillation solutions
will leave a clear footprint in the annual variation of the event
rates, with which they can be distinguished from MSW-solutions. 
Additionally, in case of the SLOW-solutions the
high energy bins of the recoil-electron energy spectrum will be 
significantly depleted compared to the other energy bins and a day-night
asymmetry of about 10\% is expected, which only can be reproduced by
the 2-flavor LOW-solution.

In this work it was investigated whether the 2-flavor solutions of the 
solar neutrino problem can be improved by
including all three flavors.
But the 3-flavor case cannot be examined totally
independent of the results obtained for the atmospheric neutrinos. 
The most promising solution to the atmospheric neutrino problem is
presently the oscillation  between $\nu_\mu$~and $\nu_\tau$~with the
mass-squared difference $\Delta m^2_{23}$ being approximately $10^{-3}\,{\rm
eV^2}$~\cite{Atm_99}. This demands
that the sum or difference of $\Delta m^2_{12}$ and $\Delta m^2_{13}$
of the 3-flavor solutions for the solar neutrino problem is about 
$10^{-3}\,{\rm eV^2}$. However, none of the presented
3-flavor solutions fulfills this condition, which implies that
either 3-flavor solutions are excluded favoring the
2-flavor LMA-solution ($\Theta_{13}=0$) or that the
origin of the atmospheric neutrino problem are not 
$\nu_\mu$-$\nu_\tau$--oscillations.  

But analyzing the atmospheric neutrino deficit  using 3-flavor neutrino
oscillations it has been 
found that with $\Delta m^2_{23}\approx10^{-4}\,{\rm eV^2}$
solutions exist,
if $\Delta m^2_{12}$ and $\sin^22\Theta_{12}$ are in the vicinity of
the SMA-solution and $4^\circ\lesssim\Theta_{13}\lesssim22^\circ$
(Fig.~2(d) in \cite{Sak99}). Indeed, the SLMA-solution fulfills this
condition as its SMA-branch ($\nu_1\!\leftrightarrow\!\nu_2$) is close to the
2-flavor SMA-solution 
and the best-fit parameters of the LMA-branch are $\Delta
m^2_{13}=2.5\times10^{-4}$ and $\Theta_{13}=19^\circ$
(cf.~Table~\ref{bestmsw3}). This solution is close to the best
3-flavor solution found recently in \cite{Fog00}, where the 825-days
data of Super-Kamiokande have been employed and where a priori
$\Delta m^2_{13}\gg10^{-4}\,{\rm eV^2}$ has been assumed. The solution in
\cite{Fog00} represents not the optimal fit to the data in the vicinity of the
SLMA-solution, as with the best-fit value for $\Delta
m^2_{13}$ being $2.5\times10^{-4}\,{\rm eV^2}$ certainly the conditions
assumed therein are not fulfilled. Therefore, a fully consistent
3-flavor analysis of the 
solar and atmospheric neutrino data should be performed to find the
most probable mixing parameters of the neutrinos.

An additional candidate, which appears to be consistent with the
atmospheric neutrino 
results, is the LVO-solution, the LMA-branch of which
($\nu_1\!\leftrightarrow\nu_3$) has similar
parameters as the SLMA-solution (Table~\ref{bestmsw3}). However, since
the oscillation 
length of the VO-branch in LVO is much larger than the Earth diameter,
$\nu_1\!\leftrightarrow\!\nu_2$-oscillations have no effect on atmospheric
neutrinos. Hence, the pure 2-flavor analysis can be performed for the
atmospheric neutrino oscillations favoring $\Delta m_{23}^2=\Delta
m^2_{13}\approx 10^{-3}\,{\rm eV^2}$.

A further possible solution to the atmospheric neutrino problem may be
the oscillation into a sterile neutrino, which is neglected in this
work. So, it cannot be excluded that 3-flavor oscillation
solutions with a sterile and an active neutrino-branch exist, which
solve the solar and atmospheric neutrino problem, too.

In the 2-flavor solutions LMA, SMA, LOW, and VO it is implicitly
assumed that the mixing into the third flavor is
negligible. Thus, these 
solutions  are not coupled to the atmospheric neutrino problem, as
they can easily be taken as $\nu_e$-$\nu_\mu$--oscillations, where no mixing
between $\nu_e$ and $\nu_\tau$ appears. 
If however the results of the LSND-col\-lab\-o\-ra\-tion~\cite{LSND98}, which
claimed to have detected $\nu_e$-$\nu_\mu$--os\-cil\-la\-tions 
with $\Delta m^2_{12}\gtrsim 10^{-2}\,{\rm eV}^2$, are confirmed by
KARMEN2~\cite{Kar99}, it will be difficult to solve the solar and
atmospheric neutrino anomaly simultaneously by the oscillation of
three neutrino flavors. A 4-neutrino scheme has been proposed  
(one sterile and three active neutrinos) to cover the three different
mass scales involved by the LSND-result, atmospheric and solar
neutrino problems~\cite{Val98}.  
In any case, identifying the solution to the solar neutrino
puzzle by future data of forthcoming and presently operating
solar-neutrino detectors may also constrain the solutions for the 
atmospheric neutrino problem or vice versa. 

The possibility that 2-flavor neutrino oscillations are
responsible for the solar neutrino puzzle appears to be supported by
the relatively weak improvement of the fit by increasing the parameter
space from 2 (two flavors) to 4 dimensions (three
flavors). Indeed, only with the SVO-solution a somewhat better fit to
the data could be achieved compared to the LMA-solution. This can be taken as a
hint, that the errors in the data of Super-Kamiokande are greater then
assumed,  perhaps still unknown systematic errors are
apparent, or that the statistics is still insufficient.
Moreover, correlations between the different
data sets of Super-Kamiokande probably are important for a
more accurate and reliable analysis of the solar neutrino data.

Anyway, if the explanation of the atmospheric neutrino deficit is,
despite the presently strong evidence,
found not to be neutrino oscillations, 3-flavor oscillations are
allowed in any mass range, implying that the SVO-solution leads to
the best fit to the data. Remarkably, SVO has been found to
be the favored solution also when taking the previous 504- or
708-days data sets of Super-Kamiokande~\cite{PhD}. Hence, SVO is
together with the 
2-flavor LMA-solution one of the stablest explanation for the solar
neutrino problem.

\acknowledgments
I am grateful to Achim Weiss and Wolfgang Hillebrandt for
their inspiring comments and suggestions. Furthermore, the support of
Georg Raffelt 
in particle-physics questions is acknowledged. I also want to thank
J{\o}rgen Christensen-Dalsgaard for his help in further improving the
solar model. Moreover, I am thankful to Sarbani Basu for providing me
the seismic model, to Scilla Degl'Innocenti for the seismic error data,
to John N.~Bahcall for the detector response functions and other
useful quantities of SNO and to Lothar Oberauer for the data about the
Borexino properties.
This work was supported by the ``Sonderforschungsbereich 375-95
f\"ur Astroteilchenphysik'' der Deutschen Forschungsgemeinschaft. 

\appendix
\section*{{From} the Initial Neutrino State to the Event Rates}
To obtain event rates for the various detectors a variety of averages
and integrals have to be evaluated. In this section briefly a
numerical scheme is described, how accurate predictions for the event
rates can be calculated numerically.

\begin{figure}[b]
\hbox to\hsize{\hss\epsfxsize=8.6cm\epsfbox{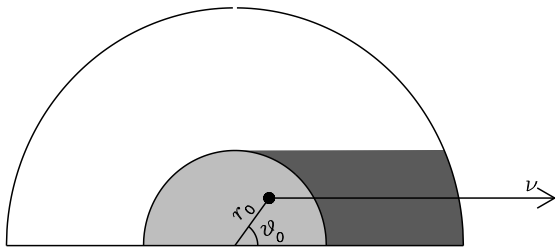}\hss}
\caption{Illustrating the definition of $r_0$ and $\vartheta_0$. The
light shaded region reflects the region of neutrino production, the
dark shaded is covered by about 50 different paths (see text) with 
direction to the Earth (arrow labeled $\nu$).\label{sunpath}}
\end{figure}

For a certain set of parameters $\Delta m^2_{12}$, $\Delta m^2_{13}$,
$\sin^2 2\Theta_{12}$, and $\sin^2 2\Theta_{13}$ the evolution of an
initially pure electron-neutrino from the solar interior to the
terrestrial detectors has to be evaluated. Therefore, 
\begin{equation} \label{neuint}
|\nu_i(r)\rangle = e^{-i\tilde{H}r}|\nu_i^e\rangle
\end{equation}
must be integrated (cf.~Eq.~\ref{schroe}). $|\nu_i^e\rangle$
denotes the initial electron-neutrino state given by 
\mbox{Eq.~(\ref{electron})}. The evolution starts at  
a point inside the Sun given by $r_0$ and $\vartheta_0$ (see
Fig.~\ref{sunpath}). 
The neutrino path has then to be followed through Sun, 
space and, if necessary, Earth arriving at the detector at the day
$D$ under the 
zenith angle $\zeta$.
{From} the final state $|\nu_i^{\rm f}\rangle$ the electron-neutrino
survival probability $P_{e\rightarrow e}(\Delta m^2_{12}/E, \Delta m^2_{13}/E, 
\sin^2 2\Theta_{12},\sin^2 2\Theta_{13}, r_0, \vartheta_0, D, \zeta)$  is derived using the
relation
\begin{equation} \label{probdef}
P_{e\rightarrow e} = \left|\langle \nu_i^{\rm f}|\nu_i^e\rangle\right|^2.
\end{equation}
Since none of the detectors is able to discriminate between $\nu_\mu$~and
$\nu_\tau$, the contribution of these neutrino families is simply given by
$1-P_{e\rightarrow e}$. 

First, the neutrino paths through the Sun are considered. Since
$N_{\rm e}$, and thus also the Hamiltonian
$\tilde{\cal H}$, is a function of $r$  the 
integration of \mbox{Eq.~(\ref{neuint})} is performed piecewise.
For this purpose, the interior of a solar model is 
subdivided into a grid equally spaced in electron density (about
500 grid points).  
By approximating the electron density between two neighboring 
grid point linearly, \mbox{Eq.~(\ref{neuint})} can be solved
analytically. Thereby the neutrino state can be
followed through the whole solar interior. Furthermore, the
accuracy is improved by increasing the grid density near the resonance
(Eq.~\ref{reson}).

Most of the neutrinos emerge between $0<r_0\lesssim 0.4 R_\odot$
and $0\le\vartheta_0\le\pi$. If the
MSW-resonance lies within this region, it is crucial to carefully 
calculate the paths for neutrinos with different $r_0$, as the more
centrally produced neutrinos may undergo a MSW-conversion, while the
outermost are hardly influenced by the matter. Paths with 
$\vartheta_0>\pi/2$ may even twice cross the MSW-resonance region.
The contribution of a neutrino emerging at an angle $\vartheta_0$ and
radius $r_0$ to the total flux originating in this shell is
given by \[
\Omega(\vartheta_0) = 0.5\sin\left(2\vartheta_0\right){\rm d}\vartheta_0\]
 for
$0\le\vartheta_0\le\pi/2$. {From} the symmetry around the plane
$\vartheta_0=\pi/2$ it follows that
$\Omega(\vartheta_0)=\Omega(\pi-\vartheta_0)$. 

Since all neutrinos cross
the shell $0.4 R_\odot\lesssim r_0 \le R_\odot$ a set of paths
through this area is established (generally about 50 equally spaced in
$\sin\vartheta_0$) on which 
$\exp({-i\tilde{\cal H}r})$ is calculated numerically by adding up the
solutions in each region linearized in $N_{\rm e}$. Thus, a set of matrices is
obtained 
${\cal S}_\odot(\sin\vartheta_0)$ which describes the evolution from
$r \approx 0.4 R_\odot$ to the solar surface. By this means the
neutrino state $|\nu_i^\odot(r_0,\vartheta_0)\rangle$ is evaluated by first
integrating \mbox{Eq.~(\ref{neuint})} until $r\approx0.4 R_\odot$ and then
using that matrix ${\cal S}_\odot(\sin\vartheta_0)$, which was calculated for a
path closest to the actual path of $|\nu_i(r_0,\vartheta_0)\rangle$.
It turns out that calculating solely two paths
($\vartheta_0 = 0$ and $\pi$)  yields mostly already acceptable results.
In this case, of course, only one matrix
$S_\odot(0)$ is needed.

Now, the state $|\nu_i^\odot(r_0,\vartheta_0)\rangle$ is followed
until the surface of the Earth using the equation for the vacuum
oscillations (Eq.~\ref{vacsol}). Since the Earth orbit is
eccentric, different days $D$ during a year are
discriminated (about 20 from perihelion to aphelion). 
In a final step the rotation of the Earth and thus the different
paths of a neutrino through the Earth's interior during one day are
considered. Here a procedure similar to the shell outside the fusion
region of the Sun is applied: A set of about
30 matrices ${\cal S}_{E}(\sin\zeta)$ is calculated integrating
$\exp({-i\tilde{\cal H}r})$ through the Earth. 

By applying finally \mbox{Eq.~(\ref{probdef})} the survival probability $P_{\rm
e\rightarrow e}(r_0, \vartheta_0, D, \zeta)$ for a certain set of 
parameters $\Delta m^2_{12}/E$, $\Delta m^2_{13}/E$, $\sin^2
2\Theta_{12}$, $\sin^2 2\Theta_{13}$
is obtained. This is the basic quantity from which the final event
rates are calculated. 

Since the experiments provide data binned in certain periods
(e.g.~day, night, annual seasons or a whole year) resp.~in certain
ranges of the zenith angle, $P_{
e\rightarrow e}(r_0, \vartheta_0, D, h)$ first has to be integrated using the
zenith-angle weights $W_\lambda(\zeta, D)$ (see e.g.~\cite{BahK97}) for
each period $D$ at the detector latitude $\lambda$.

After that, $P_\lambda^k(r_0,\vartheta_0)$ is folded with the
radial distribution $\Phi_{\rm r}(r_0)$ of the neutrino sources
inside the Sun and the angle weights
$\Omega$ discriminating the different neutrino types ($pp$, $^7$Be,
$^8$B etc., denoted by the superscript $f$). The event rates in the detectors are obtained by finally
folding $P_\lambda^f$ with the energy spectrum of each neutrino type 
$\Phi_{\rm E}^f$ and the detector response function $\Xi_i(E)$, and
summing up the contribution of all neutrino types. This yields 
\begin{equation}\label{GaEv}
N_i = \sum_f \int   P_\lambda^f\, \Xi_i(E')\, \Phi_{\rm E}^f(E')\,
{\rm d} E',
\end{equation}
where $i$ is either Ga or Cl.
The detector response function for the Homestake detector can be found e.g.
in~\cite{Bah96} and for the Ga-detectors in \cite{Bah97a}.

In Super-Kamiokande, Borexino and SNO the energy spectrum of the
recoil electrons (resp.~deuterium atoms) can be measured at least
within certain energy 
bins. Furthermore, SNO is able to discriminate CC- and NC-events, while
Super-Kamiokande and Borexino measure the total number of events arisen by any
neutrino flavor. 
According to \cite{Bah97b} the response function for these type
of experiments with the recoil-electron energy between $T_0$ and $T_1$
can be described by
\begin{eqnarray}\nonumber
\Xi_\Upsilon(E, T_0, T_1) &  = &\\ \label{cross}
&&\hspace{-8mm}\int_{T_0}^{T_1} {\rm d} T \int {\rm d} T' 
R(T, T') \frac{{\rm d} \sigma_\Upsilon(E, T')}{{\rm d} T'}, 
\end{eqnarray}
where with
\[
R(T,T') =  \frac{1}{\sqrt{2\pi}\Delta_{\rm
T'}}\,\exp\left(-\frac{(T'-T+\delta)^2}{2\Delta_{\rm T'}^2}\right) 
\]
the energy resolution of the detector is included. Therein $\delta$
accounts for a possible uncertainty in the absolute energy calibration
and $\Delta_{\rm T}$ is the energy resolution width following the photon
statistics which yields
\[\Delta_{\rm T}=\Delta_{10}\sqrt{\frac{T}{\rm 10\, MeV}}\]
with $\Delta_{10}$ being the energy resolution at 10\,MeV. The
parameters $\Delta_{10}$ and $\delta$ for Super-Kamiokande, SNO and Borexino
are summarized in Table~\ref{detpar}.
\begin{table}[ht]
\caption{Energy resolution parameters used in the present work. The
values for Super-Kamiokande and SNO are taken from \protect\cite{Bah97b}. For
Borexino  $\Delta_1$ the energy resolution width at 1\,MeV
provided by L.~Oberauer (private communication) is quoted.\label{detpar}}
\center
\begin{tabular}{cccc}
 & Super-K & SNO & Borexino \\ \hline 
$\Delta_{10}$  &  1.6\,MeV   & 1.1\,MeV  & 47\,keV \\
$\delta$       &  10\, keV   & 100\,keV & --- \\ 
\end{tabular}
\end{table}

The cross sections $\frac{{\rm d} \sigma_\Upsilon(E, T)}{{\rm d} T}$ for
Super-Kamiokande and Borexino were calculated following the description
in~\cite{Bah95}. $\Upsilon$ denotes either $\nu_e$ and $\nu_\mu$
(or equivalent $\nu_\tau$) for
the interaction of these neutrino flavors with Super-Kamiokande and Borexino
or CC resp.~NC for the reaction type in the SNO
detector.  
With $\Xi_\Upsilon(E, T_0, T_1)$ given by \mbox{Eq.~(\ref{cross})} the event rates
in Borexino resp.~Super-Kamiokande are obtained from
\begin{eqnarray} \nonumber
N_i^k(T_0, T_1) & = & \sum_f \int  \Bigl[ (1-P_\lambda^{k,
f})\,\Xi_{\nu_{\!\mu}}(E', T_0, T_1)  \\
&+& P_\lambda^{k, f}\,\Xi_{\nu_{\rm e}}(E', T_0, T_1) \Bigr] 
\;\Phi_{\rm E}^{\rm  f}(E')\, {\rm d} E',\label{skrate}
\end{eqnarray}
while the number of CC- and NC-events in SNO is defined analogous to the
Gallium and Chlorine detector (Eq.~\ref{GaEv}). With the threshold energy
being greater than 5\,MeV only the sum over
$^8$B- and $hep$-neutrinos has to be performed in Super-Kamiokande and SNO.

\newcommand{\singlet}[1]{#1}

\end{document}